\documentclass[%
 reprint,
superscriptaddress,
 amsmath,amssymb,
 aps,
pra,
]{revtex4-2}

\usepackage{graphicx}
\usepackage{dcolumn}
\usepackage{bm}

\begin{document}


\title{Transfer of nonlocality and entanglement of an open three-qubit W state in the background of dilaton black hole}

\author{Chun-yao Liu}
\affiliation{%
 College of Physics, Guizhou University, Guiyang 550025, China
}%
\affiliation{%
School of Physics and Electronic Science, Guizhou Normal University, Guiyang 550001, China
}%
 
\author{Zheng-wen Long}%
 \email{zwlong@gzu.edu.cn (corresponding author)}
\affiliation{%
College of Physics, Guizhou University, Guiyang 550025, China
}%


\author{Qi-liang He}
 \email{heliang005@163.com}
\affiliation{%
School of Physics and Electronic Science, Guizhou Normal University, Guiyang 550001, China
}%



\begin{abstract}

Constrained by the complexity of theoretical calculations, current research on genuine tripartite nonlocality (GTN) within the relativistic framework concentrates mainly on Greenberger–Horne–Zeilinger-like states, with few studies addressing W states or even general tripartite states. In this paper, we apply numerical methods to investigate how environmental decoherence and spacetime dilaton influence GTN and genuine tripartite entanglement (GTE) of W states. Our results show that GTN in the physically accessible region displays a ``sudden death'' phenomenon and that sufficiently strong decoherence completely destroys GTN. By contrast, GTE in the physically accessible region initially remains unchanged and then decays only when the dilaton parameter becomes large. Notably, the GTN and GTE in the physically accessible region can be enhanced by adjusting the decoherence parameter. Furthermore, we also find that the GTN in the physically inaccessible region cannot be generated, whereas the GTE will be produced there. This implies that GTE can cross the event horizon of a black hole and realize the redistribution of quantum entanglement. Finally, we further discuss whether the GTN can be transferred to the bipartite subsystem of the system.
\end{abstract}

\maketitle


\section{\label{sec:level1}Introduction}
Quantum nonlocality, raised by Einstein, Podolsky, and Rosen in 1935 \textsuperscript{\cite{d1}},has been extensively studied and plays a significant role in quantum information tasks \textsuperscript{\cite{3x1,3x2,3x3,3x4}}.In 1964, Bell\textsuperscript{\cite{x2}}demonstrated that any physical theory adhering to the principle of local realism conflicts with the predictions of quantum theory, which gives rise to the concept of quantum Bell nonlocality. Meanwhile, Bell also proposes the Bell inequality, which serves to determine the existence of quantum non-locality. For tripartite quantum systems, some recent experiments\textsuperscript{\cite{3x9,3x10}}have shown that the genuine tripartite nonlocality (GTN) is the strongest form of three-body correlation , which can be quantified by the violation of the Svetlichny inequality\textsuperscript{\cite{3x15}}. Quantum entanglement, first introduced by Schr¨odingeret et al.\textsuperscript{\cite{jc1}}, is one of the most counterintuitive and fundamental features of quantum mechanics. It serves as a crucial physical resource, playing a central role in various quantum information processing tasks \textsuperscript{\cite{4jc1,4jc2,4jc3,4jc4}}, including quantum teleportation\textsuperscript{\cite{4jc1,4jc2,4jc3,4jc4,4jc5}} , quantum computing\textsuperscript{\cite{4jc6,4jc7,4jc8}} , quantum cryptography \textsuperscript{\cite{4jc9,4jc10}}, dense coding \textsuperscript{\cite{4jc11}}, and quantum communication \textsuperscript{\cite{4jc15,4jc16,4jc17}}. In contrast to bipartite entanglement, several studies have revealed that genuine tripartite entanglement (GTE) exhibits distinct advantages in quantum information protocols, for instance, the exponential speedup of quantum algorithms over classical computation hinges on the utilization of GTE\textsuperscript{\cite{x27,x28}}. Quantum entanglement and nonlocality are the foundations of quantum information science and may inspire future communication technologies for space exploration.

On the other hand, as described by Einstein’s general relativity, spacetime in our Universe is intrinsically curved and generically non-inertial. Therefore, the study of relativistic effects on quantum information tasks has attracted considerable attention. In particular, recent experiments have observed gravitation-induced quantum decoherence originating from Earth’s gravitational field. This critical advance confirms that research into quantum information processing in gravitational backgrounds has formally entered an era of precision experimental tests. In the field of relativistic quantum information, the application of quantum information theory and the effective execution of quantum information tasks within the framework of black holes represent important research directions. Black holes, first predicted by Einstein's general relativity, are astrophysical objects formed via the gravitational collapse of massive stars. Up to now, a growing number of astronomical experiments have provided both direct and indirect evidence confirming the existence of black holes\textsuperscript{\cite{epjc16,epjc17,epjc18}}, such as the first image of a black hole released by the Event Horizon Telescope in 2019\textsuperscript{\cite{epjc17}}. 

Aiming to investigate how black hole effects influence quantum resources, and to develop quantum techniques for probing the fundamental information of black holes, scientists have carried out extensive research in relativistic quantum information near the event horizon of black holes\textsuperscript{\cite{epjc25,epjc26,epjc27,epjc28,epjc30,epjc42,epjc45}}. However, limited by the complexity of theoretical calculations, the aforementioned studies of GTN near black hole event horizons have predominantly focused on Greenberger–Horne–Zeilinger-like states, while investigations involving W states or even general three-qubit states remain unreported. Motivated by this research gap, the present work systematically studies the influence of Garfinkle–Horowitz–Strominger (GHS) dilaton black hole on the GTN of W states in a decohering environment, and explores whether GTN can be redistributed across the black hole event horizon.

In this paper, we study the dynamics of GTN and GTE of W state in the background of GHS dilaton black hole under decoherence. Initially, we consider the three observers Alice, Bob, and Charlie share a tripartite entangled W state in flat Minkowski spacetime. Subsequently, both Bob and Charlie are placed near the event horizon of the black hole, while Alice stays stationary in the asymptotically flat region and is coupled to a generalized amplitude damping (GAD) decoherence channel. It is shown that the physically accessible GTN exhibits ``sudden death’’ and can be completely destroyed by decoherence when the decoherence strength $r$ is sufficiently large. In contrast, the physically accessible GTE remains constant initially and starts to decay as the dilaton parameter becomes large. Furthermore, we find that the physically accessible GTN cannot cross the event horizon of the black hole, whereas the physically accessible GTE can penetrate the horizon and achieve redistribution of quantum entanglement. In particular, the GTE in both the physically accessible and inaccessible regions can be enhanced by tuning the decoherence parameter $p$. Finally, we also explore whether the physically accessible GTN can be transferred to the bipartite subsystems of the system.

This paper is organized as follows. In Sec. II, we briefly recall the measures used in this paper to quantify the entanglement and nonlocality in two-qubit and three-qubit systems. In Sec. III, we describe the quantization of Dirac fields in GHS dilaton black hole. In Sec. IV, we study the influence of dilaton effect and decoherence on the GTN and GTE. In Sec. V, we investigate the influence dilaton effect and decoherence on the Bell nonlocality and entanglement for bipartite systems. Finally, a summary and discussion are given in Sec. VI. 

\section{PRELIMINARIES}

\subsection{Entanglement}
In 1998, Wootters proposed a widely used metric for bipartite entanglement through the study of a two spin $1/2$ particle system, and derived the exact analytical expression for the entanglement of formation of an arbitrary $2\otimes2$ quantum system \textsuperscript{\cite{q50,q51,q52}}. For a bipartite quantum system with density matrix $\rho_{AB}$, the concurrence that quantifies the degree of entanglement is defined as follows

\begin{equation}
\begin{aligned}
C(\rho_{AB})= \max\left \{0,\sqrt{\lambda_{1}}-\sqrt{\lambda_{2}}-\sqrt{\lambda_{3}}-\sqrt{\lambda_{4}}\right \},
\end{aligned}
\end{equation}
here, $\lambda_{i}$ represents the eigenvalues of the ``spin-flipped" density matrix operator$\rho \left (\sigma_{y}\otimes \sigma_{y} \right )\rho ^{\ast } \left (\sigma_{y}\otimes \sigma_{y} \right )$\textsuperscript{\cite{r}}, ordered in descending magnitude. $\rho ^{\ast }$ denotes the complex conjugate of $\rho_{AB}$, and $\sigma_{y}$ refers to the Pauli $Y$ matrix, whose matrix form is $\sigma _{y}=-i\left | 0  \right \rangle \left\langle1\right| +i\left | 1 \right \rangle \left\langle0\right|$. If the density matrix $\rho_{AB}$ of the two-qubits state above takes the $X$-type form shown below
\begin{equation}
\rho^{X}=\left( \begin{array}{cccc}
a & 0 & 0 &w \\
 0& b & z& 0\\
 0& z^{\ast }& c & 0\\
w^{\ast }& 0 &0  &d
\end{array} \right),
\end{equation}
then the entanglement concurrence simplifies to
\begin{equation}
C(\rho^{X})= 2 \max \left \{0, \left | z \right |-\sqrt{a d}, \left | w \right |-\sqrt{b c} \right \} .
\end{equation}

In contrast to the mature analytical metric for bipartite entanglement, quantifying entanglement in tripartite systems is generally complex, resulting in a limited number of efficient methods for computing genuine multipartite entanglement in quantum states. One such method is $\pi-tangle$\textsuperscript{\cite{e}}, which provides a practical approach to measure the genuine entanglement of a three-qubit system. This scheme offers a clear analytical solution, formally defined as follows:
\begin{equation}
\pi _{ABC}=\frac{1}{3} \left (\pi _{A}+\pi _{B}+\pi _{C}\right ).
\end{equation}
with
\begin{equation}
\begin{aligned}
\pi _{A}&=N^{2}_{A\left ( BC \right ) }-N^{2}_{AB}-N^{2}_{AC},\nonumber\\
\pi _{B}&=N^{2}_{B\left ( AC \right ) }-N^{2}_{BA}-N^{2}_{BC},\nonumber\\
\pi _{C}&=N^{2}_{C\left ( AB \right ) }-N^{2}_{CA}-N^{2}_{CB}.
\end{aligned}
\end{equation}
here  $one-tangle$ is defined as $N_{A\left (BC\right )}$, $N_{B\left (AC\right )}$, $N_{C\left (AB\right )}$, while $two-tangle$ is defined as$N_{AB}$, $N_{AC}$, $N_{BC}$. Their specific calculation expressions are presented as follows

\begin{equation}
\begin{aligned}
N_{\alpha }\left ( \beta \gamma  \right ) &=\left \| \rho ^{T_{\alpha }}_{\alpha \beta \gamma } \right \| -1,\left ( \alpha \beta \gamma =A,B,C \right ),\nonumber\\
N_{\alpha\beta  } &=\left \| \rho ^{T_{\alpha }}_{\alpha \beta} \right \| -1,\left ( \alpha \beta  =A,B \right ).
\end{aligned}
\end{equation}

respectively, where $\left \| M \right \| $ denotes the trace norm of a matrix $M$ and $T_{\alpha }$ is the partial transpose of $\rho _{\alpha \beta \gamma }$ or $\rho _{\alpha \beta }$ \textsuperscript{\cite{e24}}. 

\subsection{Nonlocality}
In response to the Einstein-Podolsky-Rosen (EPR) paradox, J. S. Bell put forward the well-known Bell inequality, a groundbreaking achievement to mathematically differentiate between quantum mechanics theory and local hidden variable theory. Following Bell’s foundational work, scientists have put forward many different forms of Bell's inequalities. The most renowned and widely utilized in quantum information science is the Bell-Clauser-Horne-Shimony-Holt (Bell-CHSH) inequality\textsuperscript{\cite{4f52}} for a two-qubit system, which can be expressed as:
\begin{equation}
\left | Tr\left (\rho B_{CHSH}\right )\right | \le 2,
\end{equation}
with
\begin{equation}
B_{CHSH}=\vec{a}\cdot  \vec{\sigma } \otimes \left ( \vec{b} +\vec{b^{'}} \right )\cdot  \vec{\sigma }+\vec{a^{'}}\cdot  \vec{\sigma } \otimes \left ( \vec{b} -\vec{b^{'}} \right )\cdot  \vec{\sigma },
\end{equation}
here, $\vec{a}$, $\vec{a^{'}}$, $\vec{b}$, $\vec{b^{'}}$ represent a three-dimensional unit vector. This formula describes the general calculation framework. Subsequently, Horodecki et al. revised it into the form that is now commonly used \textsuperscript{\cite{by59,by60,by63,by64}}
\begin{equation}
\max_{B_{CHSH}}Tr\left (\rho  B_{CHSH}\right )=2\sqrt{M\left ( \rho  \right )} .
\end{equation}

In the above equation, $M\left ( \rho  \right )=\max_{i<j} \left ( \mu _{i}+\mu _{j} \right )$, $\mu _{i}\left ( i=1,2,3 \right )$ denotes the eigenvalue of matrix $U=X^{T}X$. Matrix $X$ represents a three-dimensional correlation matrix constructed from the density matrix $\rho$, with its matrix elements defined as $X_{ij}=Tr[\rho \left ( \sigma _{i}\otimes \sigma _{j} \right )] $. By jointly evaluating Eq.(5) and Eq.(7), it can be rigorously confirmed that the Bell-CHSH inequality is violated if and only if $M\left ( \rho  \right ) > 1$, which signifies the presence of nonlocality in the quantum state. 

Conversely, if $M\left ( \rho  \right ) \le 1$, then the quantum state exhibits no nonlocality. In particular, consider that the two-body density matrix $\rho$ introduced above takes the form $\rho^X$ as defined in Eq.(2), the three eigenvalues of matrix $U\left ( \rho  \right ) =X^{T}_{\rho}X_{\rho}$ can be obtained as 
\begin{eqnarray}
Z_{1}&=&4\left (\left | w \right |  +\left | w \right | \right )^{2},\nonumber\\
Z_{2}&=&4\left (\left | w \right |  -\left | w \right | \right )^{2},\nonumber\\
Z_{3}&=&\left (\left | a\right | -\left | b \right | -\left |c \right | +\left | d \right | \right )^{2}.
\end{eqnarray}

Since $Z_{1}> Z_{2}$, the formula for calculating its Bell nonlocality can be simplified as:
\begin{equation}
B\left ( \rho _{x} \right )=\max\left \{ B_{1},B_{2} \right \} ,
\end{equation}
with $B_{1}=2\sqrt{Z_{1}+Z_{2}}$, $B_{2}=2\sqrt{Z_{1}+Z_{3}}$\textsuperscript{\cite{by65,by66}}.

In addition, in the tripartite scenario, the genuine tripartite nonlocality can be identified through the violation of the Svetlichny inequalities\textsuperscript{\cite{3f14}}
\begin{equation}
tr(S\rho_{abc}) \le 4,
\end{equation}
where the maximum value violating the Svetlichny inequality of the tripartite state $\rho_{abc}$ can be denoted as
\begin{equation}
S(\rho_{abc} )\equiv \max_{S} tr(S\rho_{abc} ).
\end{equation}

Equivalently, if $S(\rho_{abc})>4$, we can say that the tripartite state $\rho_{abc}$ must exhibit GTN \textsuperscript{\cite{n}}. However, for non-X-type tripartite quantum states, there is still no general analytical expression available for quantifying GTN at present. Therefore, in this paper, we employ an ``exact'' numerical method to determine whether the Svetlichny inequality is violated for arbitrary three-qubit states\textsuperscript{\cite{3x}}. 

\section{Quantization of the Dirac field in GHS dilaton black hole}

In this section, we first briefly analyze the vacuum structure of a Dirac particle in the GHS dilaton black hole. Following the approach used for the Schwarzschild spacetime\textsuperscript{\cite{L42}}, the metric for the GHS dilaton black hole takes the form\textsuperscript{\cite{L53}}
\begin{equation}
   {\mathrm{d}}s^2=-(\frac{r-2 M}{r-2 \alpha}) {\mathrm{d}}t^2+(\frac{r-2 M}{r-2 \alpha})^{-1} {\mathrm{d}}r^2+r (r-2 \alpha)   {\mathrm{d}}\Omega^2,
\end{equation}
here $M$ denotes the mass parameter of the black hole, and $\alpha$ is the dilaton parameter of the GHS black hole. The dilaton $\alpha$ and the mass M satisfy the relationship $\alpha=Q^2/2M$. The well-known Hawking temperature takes the form $T=\frac{1}{8 \pi (M-\alpha )} $ , and the thermal Fermi-Dirac distribution of particles of the GHS dilaton black hole with $T$ was reported in Refs.\cite{s69,s71}. For simplicity, we set $G=c=\hbar=\kappa_{B}=1$ in this paper \textsuperscript{\cite{m}} .

For GHS dilaton spacetime,  the Dirac equation for spinor field $\Psi$ can be expressed as\textsuperscript{\cite{i,i34,i35}}
\begin{equation}
    [\gamma^{a}  e_{a}^{\mu}(\partial_{\mu}+\Gamma _{\mu})]  \Psi = 0,
\end{equation}
where $\gamma^{a}$ denotes the Dirac matrices, the four-vectors $e_{a}^{\mu}$ represent the inverse of the tetrad $e^{a}_{\mu}$, and $\partial \mu $ stands for the spin connection coefficient. $\Gamma _{\mu}$ is the spin connection coefficient. By solving the Dirac equation under the GHS dilaton 
spacetime, we can obtain the following positive frequency outgoing solutions for the outside region ${\uppercase\expandafter{\romannumeral1}}$ and inside region ${\uppercase\expandafter{\romannumeral2}}$ near the event horizon \textsuperscript{\cite{s71,i44}}

\begin{eqnarray}
&&\Psi_{k}^{{\uppercase\expandafter{\romannumeral1}+}} =\xi e^{-i\omega u}, \nonumber\\
&&\Psi_{k}^{{\uppercase\expandafter{\romannumeral2}+}} =\xi e^{i\omega u},
\end{eqnarray}
here $k$ denotes the field mode,  $\xi$ is the our-component Dirac spinor, $\omega$ represents the monochromatic frequency of the Dirac field. The retarded time is defined as $u = t - r_{\ast}$, with $r_{\ast}$ representing the tortoise coordinate in the GHS spacetime, given by $r_{\ast} = r + 2(M-\alpha) \ln\left[\frac{r - 2(M-\alpha)}{2(M-\alpha)}\right]$. Substituting $r_{\ast}$ into the above expression yields the following form of the Dirac field near the event horizon.
\begin{eqnarray}
\Psi_{out}=\sum_{\sigma={\uppercase\expandafter{\romannumeral1}},{\uppercase\expandafter{\romannumeral2}}}\int\mathrm{d}k(a^{\sigma}_{k} \Psi^{\sigma +}_{k})
(b^{\sigma \ast}_{k} \Psi^{\sigma -}_k),
\end{eqnarray}
with $a^{\sigma}_{k}$ and $b^{\sigma \ast}_{k}$ denote the fermion annihilation and antifermion creation operators acting on the quantum state of the Dirac field in the dilaton black hole spacetime, respectively. By analyzing $\Psi_{k}^{{\uppercase\expandafter{\romannumeral1} +}}$ and $\Psi_{k}^{{\uppercase\expandafter{\romannumeral2} +}}$, we can construct a complete set of orthogonal bases. Furthermore, based on the relation between black hole coordinates and Kruskal coordinates, we can derive a complete basis for positive-energy modes as 

\begin{eqnarray}
&&\xi^{{\uppercase\expandafter{\romannumeral1}}_{+}}_{k}=e^{2(M-\alpha)\pi \omega }\Psi^{{\uppercase\expandafter{\romannumeral1} +}}_{k}+
e^{-2(M-\alpha)\pi \omega }\Psi^{{\uppercase\expandafter{\romannumeral2} -}}_{-k},\nonumber\\
&&\xi^{{\uppercase\expandafter{\romannumeral2}}_{+}}_{k}=e^{-2(M-\alpha)\pi \omega }\Psi^{{\uppercase\expandafter{\romannumeral1} -}}_{-k}+
e^{2(M-\alpha)\pi \omega }\Psi^{{\uppercase\expandafter{\romannumeral2} +}}_{k}.
\end{eqnarray}

Substituting the above basis into Eq.(15) and following the suggestion of Damoar–Ruffini\textsuperscript{\cite{s71}}, we obtain the Kruskal modes, which form a complete basis for positive-energy solutions. In addition to this, we also expand the Dirac field in terms of these Kruskal modes\textsuperscript{\cite{m}}.
\begin{eqnarray}
\Psi_{out}=&&\sum_{\sigma={\uppercase\expandafter{\romannumeral1}},{\uppercase\expandafter{\romannumeral2}}}\int\mathrm{d}k \frac{1}{\sqrt{2\cosh [4(M-\alpha )\pi \omega ]}}\nonumber\\ 
&&(c^{\sigma}_{k} \xi^{\sigma +}_{k})
d^{\sigma \ast}_{k} \xi^{{\sigma}_{-}})
\end{eqnarray}
here, $c^{\sigma}_{k}$ and $d^{\sigma \ast}_{k}$ denote the fermion annihilation and antifermion creation operators in the Kruskal vacuum that act on the exterior and interior region states, respectively. Using the Dirac field decomposition in the GHS dilaton black hole given in Eq.(15) and the Kruskal modes in Eq.(17), we derive the Bogoliubov transformations between Kruskal and dilaton modes. From consideration of the mode orthonormality conditions, it can be shown that each Kruskal annihilation operator $c^{\uppercase\expandafter{\romannumeral1}}_{k}$ mixes only with dilaton particle operators of a single frequency $\omega_{i}$, and therefore $c^{\uppercase\expandafter{\romannumeral1}}_{k}$ assumes the following form for the sector\textsuperscript{\cite{L}}.
\begin{eqnarray}
&&c^{\uppercase\expandafter{\romannumeral1}}_{k}=\cos{\theta}\cdot a^{\uppercase\expandafter{\romannumeral1}}_{k}
-\sin{\theta}\cdot   b^{{\uppercase\expandafter{\romannumeral1}}_{\ast}}_{k},\nonumber\\ 
&&\sin{\theta}=[e^{-8 \pi \omega_{i} (M-\alpha)}+1]^{-\frac{1}{2}},\nonumber\\ 
&&\cos{\theta}=[e^{8 \pi \omega_{i} (M-\alpha)}+1]^{-\frac{1}{2}}.
\end{eqnarray}

Due to the existence of physically accessible and inaccessible regions in the GHS dilaton spacetime, the ground state mode in the generalized GHS dilaton black hole coordinates can be mapped to the two-mode squeezed state in the Kruskal coordinates. After properly normalizing the state vector, the vacuum and only excited states for a given Kruskal particle mode can be expressed as follows:
\begin{eqnarray}
&&\left | 0_{k} \right \rangle _{k}^{+}=\cos{\theta}\cdot \left | 0_{k} \right \rangle _{{\uppercase\expandafter{\romannumeral1}}}^{+}\left | 0_{-k} \right \rangle _{{\uppercase\expandafter{\romannumeral2}}}^{-}
+\sin{\theta}\cdot \left | 1_{k} \right \rangle _{{\uppercase\expandafter{\romannumeral1}}}^{+}\left | 1_{-k} \right \rangle _{{\uppercase\expandafter{\romannumeral2}}}^{-},\nonumber\\ 
&&\left | 1_{k} \right \rangle _{k}^{+}= \left | 1_{k} \right \rangle _{{\uppercase\expandafter{\romannumeral1}}}^{+}\left | 0_{-k} \right \rangle _{{\uppercase\expandafter{\romannumeral2}}}^{-},
\end{eqnarray}
where $\left |n \right \rangle _{\uppercase\expandafter{\romannumeral1}}$ and $\left |n \right \rangle _{\uppercase\expandafter{\romannumeral2}}$ correspond to the orthonormal bases for the outside and inside regions of the event horizon, respectively, the superscripts $\left \{ +,- \right \} $ of the basis vectors denote the particle and antiparticle vacua. For simplicity, we set $\omega=\omega_{i}=1$.

\section{The dynamics of GTN and GTE of W state in a dilaton black hole under decoherence}

We assume observers Alice, Bob, and Charlie initially stay at the same point in flat Minkowski spacetime and share a tripartite entangled three-qubit $W$ state, which can be expressed as 
\begin{equation}
   \left |W \right \rangle=\frac{1}{\sqrt{3} }[\left | 0_{A}0_{B}1_{C} \right \rangle+
   \left | 0_{A}1_{B}0_{C} \right \rangle +\left | 1_{A}0_{B}0_{C} \right \rangle]
\end{equation}

Subsequently, we consider that Alice remains stationary in the asymptotically flat region, while Bob and Charlie fall freely into the black hole and eventually hover near the event horizon. By employing the GHS dilaton black hole quantization method and adopting the single-mode approximation described in Eq.(19), we can rewrite the W state given in Eq.(20) as

\begin{eqnarray}
   \left |W \right \rangle&=&\frac{1}{\sqrt{3} }[\cos{\theta} \left | 0_{A}0_{B_{\uppercase\expandafter{\romannumeral1}}}0_{B_{\uppercase\expandafter{\romannumeral2}}}                 1_{C_{\uppercase\expandafter{\romannumeral1}}}0_{C_{\uppercase\expandafter{\romannumeral2}}}\right \rangle\nonumber\\
   &+&\cos{\theta}\left | 0_{A}1_{B_{\uppercase\expandafter{\romannumeral1}}}0_{B_{\uppercase\expandafter{\romannumeral2}}}                 0_{C_{\uppercase\expandafter{\romannumeral1}}}0_{C_{\uppercase\expandafter{\romannumeral2}}}\right \rangle\nonumber\\
   &+&\sin{\theta}\left | 0_{A}1_{B_{\uppercase\expandafter{\romannumeral1}}}0_{B_{\uppercase\expandafter{\romannumeral2}}}                 1_{C_{\uppercase\expandafter{\romannumeral1}}}1_{C_{\uppercase\expandafter{\romannumeral2}}}\right \rangle\nonumber\\
   &+&\sin{\theta} \left | 0_{A}1_{B_{\uppercase\expandafter{\romannumeral1}}}1_{B_{\uppercase\expandafter{\romannumeral2}}}                 1_{C_{\uppercase\expandafter{\romannumeral1}}}0_{C_{\uppercase\expandafter{\romannumeral2}}}\right \rangle\nonumber\\
   &+&\cos{\theta}^{2} \left | 1_{A}0_{B_{\uppercase\expandafter{\romannumeral1}}}0_{B_{\uppercase\expandafter{\romannumeral2}}}                 0_{C_{\uppercase\expandafter{\romannumeral1}}}0_{C_{\uppercase\expandafter{\romannumeral2}}}\right \rangle\nonumber\\
   &+&\cos{\theta} \sin{\theta}\left | 1_{A}0_{B_{\uppercase\expandafter{\romannumeral1}}}0_{B_{\uppercase\expandafter{\romannumeral2}}}                 1_{C_{\uppercase\expandafter{\romannumeral1}}}1_{C_{\uppercase\expandafter{\romannumeral2}}}\right \rangle\nonumber\\
   &+&
   \cos{\theta} \sin{\theta}\left | 1_{A}1_{B_{\uppercase\expandafter{\romannumeral1}}}1_{B_{\uppercase\expandafter{\romannumeral2}}}                 0_{C_{\uppercase\expandafter{\romannumeral1}}}0_{C_{\uppercase\expandafter{\romannumeral2}}}\right \rangle\nonumber\\
   &+&\sin{\theta}^{2} \left | 1_{A}1_{B_{\uppercase\expandafter{\romannumeral1}}}1_{B_{\uppercase\expandafter{\romannumeral2}}}                 1_{C_{\uppercase\expandafter{\romannumeral1}}}1_{C_{\uppercase\expandafter{\romannumeral2}}}\right \rangle],
\end{eqnarray}

We assume Alice's detector senses only the modes $\left |n \right \rangle_{A}$ outside the event horizon, while Bob's and Charlie's detectors observe the modes $\left |n \right \rangle_{B_{{\uppercase\expandafter{\romannumeral1}}}}$, $\left |n \right \rangle_{C_{{\uppercase\expandafter{\romannumeral1}}}}$ outside the black hole event horizon. Meanwhile, the detectors for the antiparticle modes, anti-Bob and anti-Charlie, observe the modes $\left |n \right \rangle_{B_{{\uppercase\expandafter{\romannumeral2}}}}$ and $\left |n \right \rangle_{C_{{\uppercase\expandafter{\romannumeral2}}}}$ inside the event horizon.

Here, we consider that Alice remains at the same point and interacts with a dissipative environment, which can be modeled by the well-known generalized amplitude damping (GAD) channel, denoted as 
\begin{equation}
 \varepsilon _{GAD} (\rho )=\sum_{i=0}^{3} E_{i} \rho E_{i}^{\dagger },
\end{equation}
with
\begin{eqnarray}
&&E_{0}=\sqrt{p}
  \left( \begin{array}{cc}
  1 & 0 \\
 0 &\sqrt{1-r}
\end{array} \right), E_{1}=\sqrt{p}  \left( \begin{array}{cc}
  0&\sqrt{r} \\
 0 &0
\end{array} \right), \nonumber\\
&&E_{2}=\sqrt{1-p}
\left( \begin{array}{cc}
  \sqrt{1-r}&0 \\
 0 &1
\end{array} \right), \nonumber\\ &&E_{3}=\sqrt{1-p}  \left( \begin{array}{cc}
  0&0 \\
 \sqrt{r} &0
\end{array} \right),
\end{eqnarray}
where $r$ is linked to the rate of energy relaxation, and $\left \{ p,r \right \} $ depends on the temperature ${T}'$ of the environment, which can be expressed as
\begin{eqnarray}
&r&=1-e^{-\gamma t}, \nonumber\\
&\gamma& =[\frac{2}{\exp (-\frac{h \omega }{k_B {T}' } )-1}+1 ] \gamma _{0}, \nonumber\\
&p&=\frac{1}{1+{\exp (-\frac{h \omega }{k_B {T}' }) }} .
\end{eqnarray}
with $\gamma_{0}$ characterizes the energy relaxation rate, $t$ denotes the duration of storage, and $h \omega$ and $k_{B}$ are the transition energy of the quantum system and the Boltzmann constant, respectively.

By utilizing Eqs.(21) and (22), we can derive the  density matrix of system as 
\begin{eqnarray}
\left |\rho^{'} \right \rangle_{A B_{\uppercase\expandafter{\romannumeral1}} B_{\uppercase\expandafter{\romannumeral2}} C_{\uppercase\expandafter{\romannumeral1}} C_{\uppercase\expandafter{\romannumeral2}}}= \varepsilon _{GAD} (\rho )=\sum_{i=0}^{3} E_{i} \rho E_{i}^{\dagger },
\end{eqnarray}

Since the event horizon of a black hole isolates its interior region from the exterior region, neither Alice, Bob, nor Charlie can access the modes inside the event horizon. In view of this, we refer to Mode $B_{\uppercase\expandafter{\romannumeral1}}$, $ C_{\uppercase\expandafter{\romannumeral1}} $ outside the black hole event horizon as the mode of the physically accessible region, and Mode $B_{\uppercase\expandafter{\romannumeral2}}$, $ C_{\uppercase\expandafter{\romannumeral2}} $ inside the event horizon as the mode of the physically inaccessible region. Therefore, by tracing out the physically inaccessible modes in Eq.(25), we obtain the density matrix for the physically accessible region modes as 
\begin{equation}
\rho_{A B_{\uppercase\expandafter{\romannumeral1}} C_{\uppercase\expandafter{\romannumeral1}} }= \left( \begin{array}{cccccccc}
 \chi_{11} & 0 & 0 &0& 0 & 0& 0 & 0 \\
  0& \chi _{22} & \chi_{23} & 0& \chi_{25} & 0& 0 & 0\\
  0& \chi_{32} & \chi _{33} & 0& \chi _{35} & 0& 0 & 0\\
 0 & 0 &0  & \chi _{44}& 0 & \chi _{46}& \chi _{47} & 0\\
 0 & \chi _{52} &\chi_{53}  &0& \chi _{55} & 0& 0 & 0\\
 0 & 0 &0  &\chi_{64}& 0 & \chi _{66}& \chi _{67} & 0\\
 0 & 0 &0  &\chi _{74}& 0 & \chi _{76}& \chi _{77} & 0\\
0 & 0 &0  &0& 0 & 0& 0 & \chi _{88}
\end{array} \right),
\end{equation}
with

\begin{eqnarray*}
&&\chi_{11}=\frac{1}{3} \beta ^{4}pr,\nonumber\\
&&\chi_{22}=\chi_{33}=-\frac{1}{3}\beta ^{2}(-1+r+(\beta ^{2}-2)pr), \nonumber\\
&&\chi_{23}=\chi_{32}=\frac{1}{3}\beta ^{2}(1+(p-1)r),\nonumber\\
&&\chi_{25}=\chi_{35}=\chi_{52}=\chi_{53}=\frac{1}{3} \beta ^{3}\sqrt{1-r}\nonumber\\
&&\chi_{44}=\frac{1}{3}(\beta ^{2}-1)( ( (\beta ^{2}-3)p+2)r-2)\nonumber\\
&&\chi_{46}=\chi_{47}=\chi_{64}=\chi_{67}=-\frac{1}{3} \beta (\beta ^{2}-1)\sqrt{1-r},\nonumber\\
&&\chi_{55}=-\frac{1}{3} \beta ^{4}(pr-1),\nonumber\\
&&\chi_{66}=\chi_{77}=\frac{1}{3} \beta^{2}(1+r-2pr+\beta ^{2}(pr-1)),\nonumber\\
&&\chi_{67}=\chi_{76}=\frac{1}{3} \beta^{2}(1-p)r,\nonumber\\
&&\chi_{88}=\frac{1}{3} (\beta ^{2}-1)(\beta ^{2}-((\beta ^{2}-3)p+2)r-1),\nonumber\\
&&\beta=(e^{-8\pi\omega(M-\alpha)}+1)^{-\frac{1}{2}}.
\end{eqnarray*}

Using numerical methods for GTN\textsuperscript{\cite{3x}} and the $\pi$-tangle \textsuperscript{\cite{e}} approach for GTE, we can obtain the GTN and GTE in the physically accessible region. 

\begin{figure}
\begin{center}
\includegraphics[width=6cm]{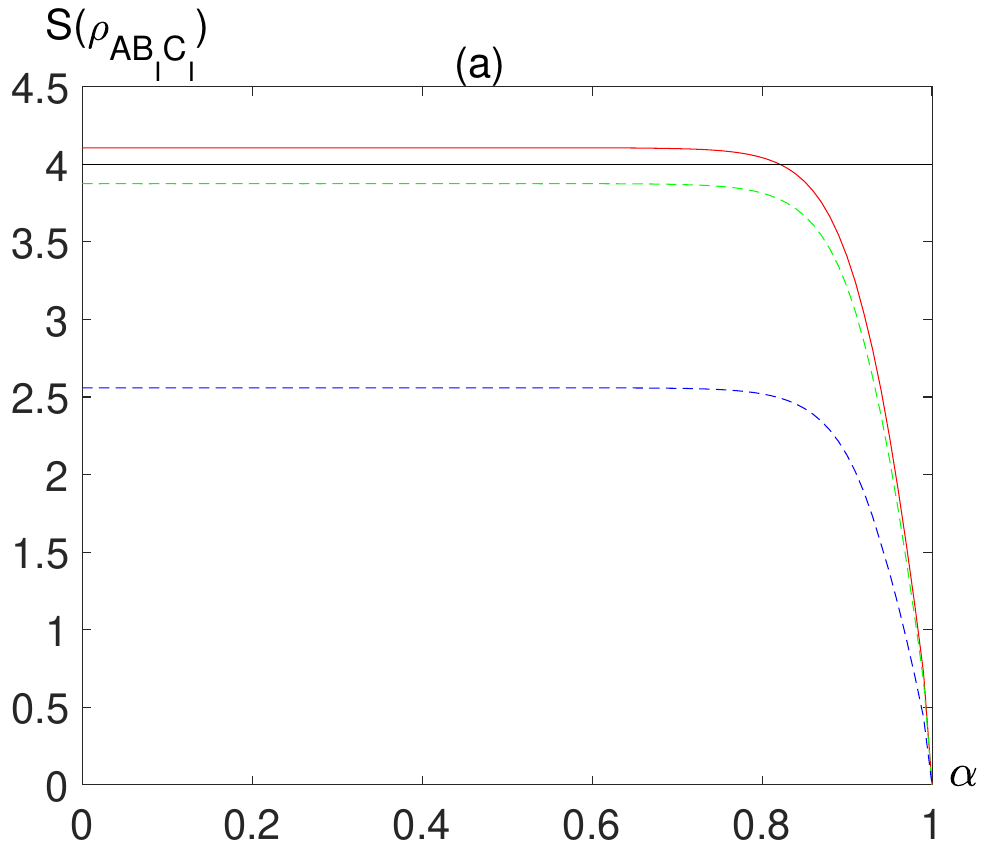}
\includegraphics[width=6cm]{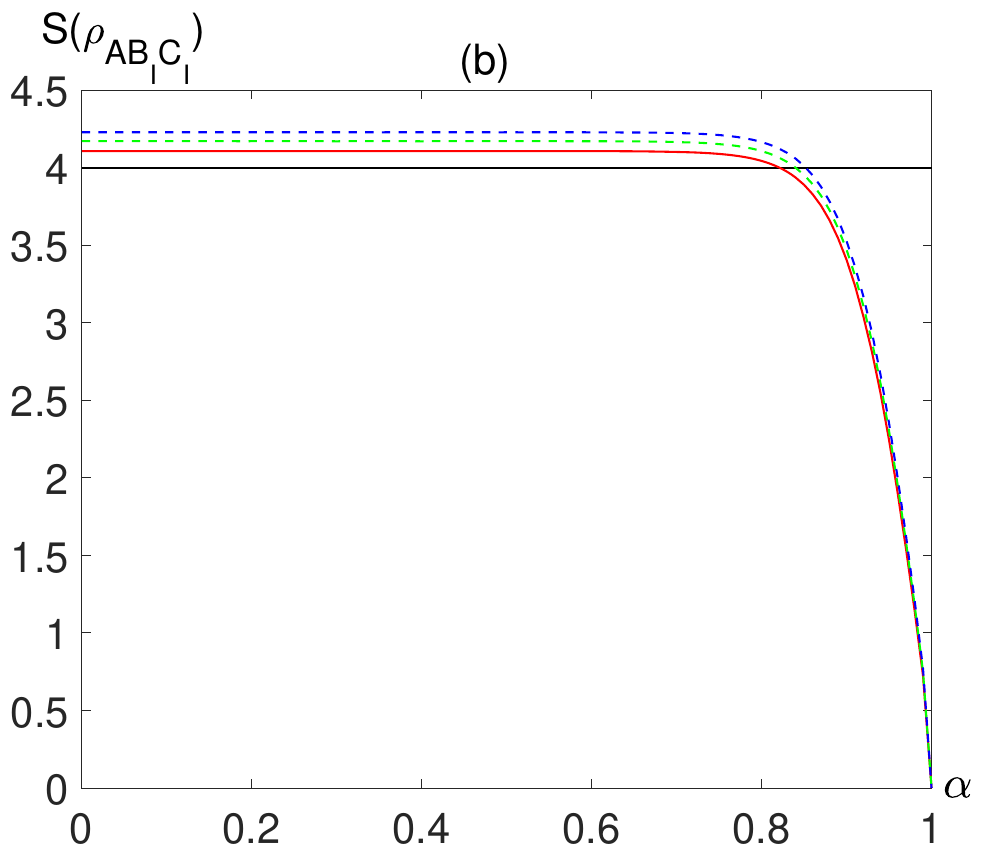}
\caption{\label{fig:fig1} (a) GTN $S(\rho_{AB_IC_I})$ of the three-qubit system as a function of the dilaton parameter $\alpha$ with $\omega=1$, $M=1$ for r=0.05, p=0.05(red solid line), r=0.1, p=0.1 (green dashed line) and r=0.5, p=0.5 (blue dashed line). (b) GTN $S(\rho_{AB_IC_I})$ of the three-qubit system as a function of the dilaton parameter $\alpha$ with $\omega=1$, $M=1$ for $r=0.05$, $p=0.05$ (red solid line), $r=0.05$, $p=0.5$ (green dashed line) and $r=0.05$, $p=0.9$ (blue dashed line).}
\end{center}
\end{figure}

In Fig.1, we show the GTN $S(\rho_{A B_{\uppercase\expandafter{\romannumeral1}} C_{\uppercase\expandafter{\romannumeral1}} })$ in the physically accessible region as a function of the dilaton parameter $\alpha$ for different decoherence strengths $r$ and decoherence parameters $p$ of the GAD channel. From the Fig.1(a), we can find that as $\alpha$ increases, $S(\rho_{A B_{\uppercase\expandafter{\romannumeral1}} C_{\uppercase\expandafter{\romannumeral1}} })$ initially exceeds $4$ and then falls below $4$. This behavior reveals that GTN exhibits "sudden death" as the dilaton parameter $\alpha$ varies, which means that the dilaton effect of the black hole destroys the GTN in the physically accessible region shared by Alice, Bob, and Charlie. By Comparing the three curves in the Fig.1(a), it is evident that both the initial magnitude of GTN and the critical dilaton parameter at which sudden death occurs will decrease as the decoherence strength $r$ increases. In particular, when the decoherence strength $r$ is relatively large, the GTN remains consistently below $4$. This indicates that environmental decoherence completely destroys the GTN in the physically accessible regions. In addition, it is clear from Fig.1(b) that as the decoherence parameter $p$ increases, the value of the GTN $S(\rho_{A B_{\uppercase\expandafter{\romannumeral1}} C_{\uppercase\expandafter{\romannumeral1}} })$ in the physically accessible region be increased and the critical $\alpha$ at which ``sudden death'' occurs shifts is shifted to a larger value. This result implies that by adjusting the decoherence parameter $p$, we can enhance the GTN in the physically accessible region and simultaneously delay its sudden death, which is beneficial for the implementation of quantum information tasks based on quantum nonlocality.

\begin{figure}
\begin{center}
\includegraphics[width=6cm]{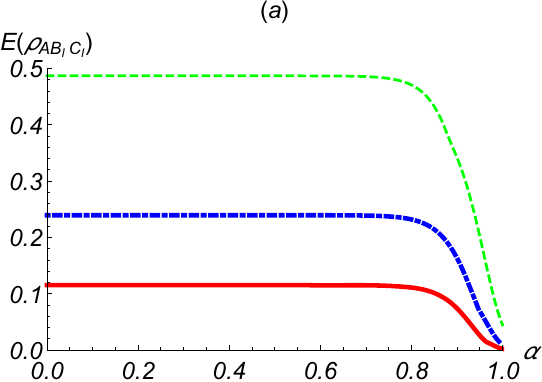}
\includegraphics[width=6cm]{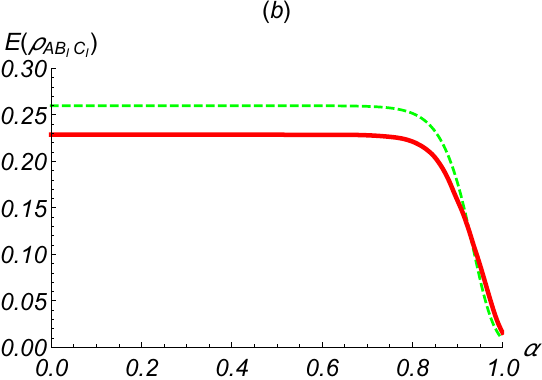}
\caption{\label{fig:fig2} 
(a)  GTE $E(\rho_{AB_IC_I})$ of the three-qubit system as a function of the dilaton parameter $\alpha$ with $\omega=1$, $M=1$ for $r=0.1$, $p=0.5$ (green dashed line), $r=0.5$, $p=0.5$ (blue dashed line) and $r=0.7$, $p=0.5$ (red solid line). (b) GTE $E(\rho_{AB_IC_I})$ of the three-qubit system as a function of the dilaton parameter $\alpha$ with $\omega=1$, $M=1$ for $r=0.5$, $p=0.05$ (green dashed line), and $r=0.5$, $p=0.9$ (red solid line).}
\end{center}
\end{figure}

In Fig.2, we display how the GTE $E(\rho_{A B_{\uppercase\expandafter{\romannumeral1}} C_{\uppercase\expandafter{\romannumeral1}} })$ in the physically accessible region varies with the dilaton parameter $\alpha$ for different decoherence strengths $r$ and decoherence parameters $p$. It is demonstrated in the Fig. 2(a) that as dilaton parameter $\alpha$ increases, $E(\rho_{A B_{\uppercase\expandafter{\romannumeral1}} C_{\uppercase\expandafter{\romannumeral1}} })$ initially remains nearly constant and then begins to decay once $\alpha$ becomes relatively large. This implies that the dilaton effect of the black hole has a negative impact on the GTE $E(\rho_{A B_{\uppercase\expandafter{\romannumeral1}} C_{\uppercase\expandafter{\romannumeral1}} })$. By comparing the three curves in the figure, it can be clearly seen that the GTE $E(\rho_{A B_{\uppercase\expandafter{\romannumeral1}} C_{\uppercase\expandafter{\romannumeral1}} })$ decreases with the increase of the decoherence strength $r$. In particular, we observe that the GTE does not undergo “sudden death”, which indicates that the decoherence does not completely destroy the GTE $E(\rho_{A B_{\uppercase\expandafter{\romannumeral1}} C_{\uppercase\expandafter{\romannumeral1}} })$. This result is different from the case of GTN in Fig. 1(a), demonstrating that GTE has stronger robustness against noise than GTN. This may provide a reference for selecting appropriate quantum resources for quantum information tasks in the presence of decoherence. From Fig.2(b), we can see that as the decoherence parameter $p$ decreases, the GTE $E(\rho_{A B_{\uppercase\expandafter{\romannumeral1}} C_{\uppercase\expandafter{\romannumeral1}} })$ in the physically accessible region increases, which means that the GTE can be enhanced by adjusting the decoherence parameter $p$. The physical interpretation of this result is that in the GAD channel, the decoherence parameter $p$ corresponds to the relative probability $\frac{p}{1-p}$ of excitation loss versus excitation gain  in the quantum system under the GAD channel. A smaller $p$ therefore reduces the probability of excitation loss from the quantum system to the environment. Consequently, decreasing the decoherence parameter $p$ increases the GTE in the physically accessible region of the quantum system. In addition, by comparing Fig.1(b) and Fig.2(b), we can clearly see that the influence of the decoherence parameter $p$ on GTN and GTE is different. Specifically, GTN increases as decoherence parameter $p$ increases, whereas GTE increases as decoherence parameter $p$ decreases. These results demonstrate that GTN and GTE do not exhibit simple monotonic relationship, namely, an increase in GTE does not necessarily produce a corresponding increase in GTN, and in some cases, GTN can even decrease \textsuperscript{\cite{hb1,hb2}}.

To further characterize the evolution of GTN and GTE in other tripartite subsystems, we proceed to analyze these quantum resources in the tripartite subsystem involving physically inaccessible region inside the event horizon.  Tracing out modes $C_{\uppercase\expandafter{\romannumeral1}}$ and $C_{\uppercase\expandafter{\romannumeral2}}$ from Eq.(25), we obtain the reduced density matrix $\rho_{A B_{\uppercase\expandafter{\romannumeral1}} B_{\uppercase\expandafter{\romannumeral2}} }$, whose explicit form is given by
\begin{equation}
\rho_{A B_{\uppercase\expandafter{\romannumeral1}} B_{\uppercase\expandafter{\romannumeral2}} }=
 \left( \begin{array}{cccccccc}
 \nu_{11} & 0 & 0 &\nu_{14}& 0 & 0& 0 & 0 \\
  0& 0& 0 & 0& 0 & 0& 0 & 0\\
  0& 0 & \nu_{33} & 0&\nu_{35} & 0& 0 & \nu_{38}\\
 \nu_{41} & 0 &0  & \nu_{44}& 0 & 0& 0 & 0\\
 0 & 0 &\nu_{53}  &0& \nu_{55} & 0& 0 & \nu_{58}\\
 0 & 0 &0  &0& 0 & 0& 0 & 0\\
 0 &0 &0  &0& 0 & 0& \nu_{77} & 0\\
 0 & 0 &\nu_{83} &0& \nu_{85} & 0& 0 & \nu_{88}
\end{array} \right),
\end{equation}
with
\begin{eqnarray*}
&&\nu_{11}=\frac{1}{3}\beta ^{2}(1+(2p-1)r),\nonumber\\
&&\nu_{14}=\nu_{41}=\frac{1}{3} \beta \sqrt{1-\beta ^{2}}(1+(2p-1)r),\nonumber\\
&&\nu_{33}=\frac{1}{3}(1+(p-1)r),\nonumber\\
&&\nu_{35}=\nu_{53}=\frac{1}{3} \beta \sqrt{1-r},\nonumber\\
&&\nu_{38}=\nu_{83}=\frac{1}{3} \sqrt{1-\beta^{2}} \sqrt{1-r},\nonumber\\
&&\nu_{44}=\frac{1}{3}(1-\beta^{2})(1+(2p-1)r),\nonumber\\
&&\nu_{53}=\frac{1}{3} \beta \sqrt{1-r},\nonumber\\
&&\nu_{55}=\frac{1}{3} \beta^{2} (1+r-2pr),\nonumber\\
&&\nu_{58}=\nu_{85}=\frac{1}{3} \beta \sqrt{1-\beta^{2}} (1+r-2pr), \nonumber\\
&&\nu_{77}=\frac{1}{3} (1-p)r, \nonumber\\
&&\nu_{88}=\frac{1}{3}(\beta^{2}-1)((2p-1)r-1), \nonumber\\
&&\beta=(e^{-8\pi\omega(M-\alpha)}+1)^{-\frac{1}{2}}.
\end{eqnarray*}

Following the same method as described above, we can obtain the GTN and GTE corresponding to the density matrix $\rho_{A B_{\uppercase\expandafter{\romannumeral1}} B_{\uppercase\expandafter{\romannumeral2}} }$. Based on the exchange symmetry between Bob and Charlie, we find that GTN and GTE of the tripartite subsystem $\rho_{A C_{\uppercase\expandafter{\romannumeral1}} C_{\uppercase\expandafter{\romannumeral2}} }$ are equal to those of the subsystem $\rho_{A B_{\uppercase\expandafter{\romannumeral1}} B_{\uppercase\expandafter{\romannumeral2}} }$. Therefore, it suffices to analyze the GTN and GTE of the density matrix $\rho_{A B_{\uppercase\expandafter{\romannumeral1}} B_{\uppercase\expandafter{\romannumeral2}} }$.

\begin{figure}
\begin{center}
\includegraphics[width=6cm]{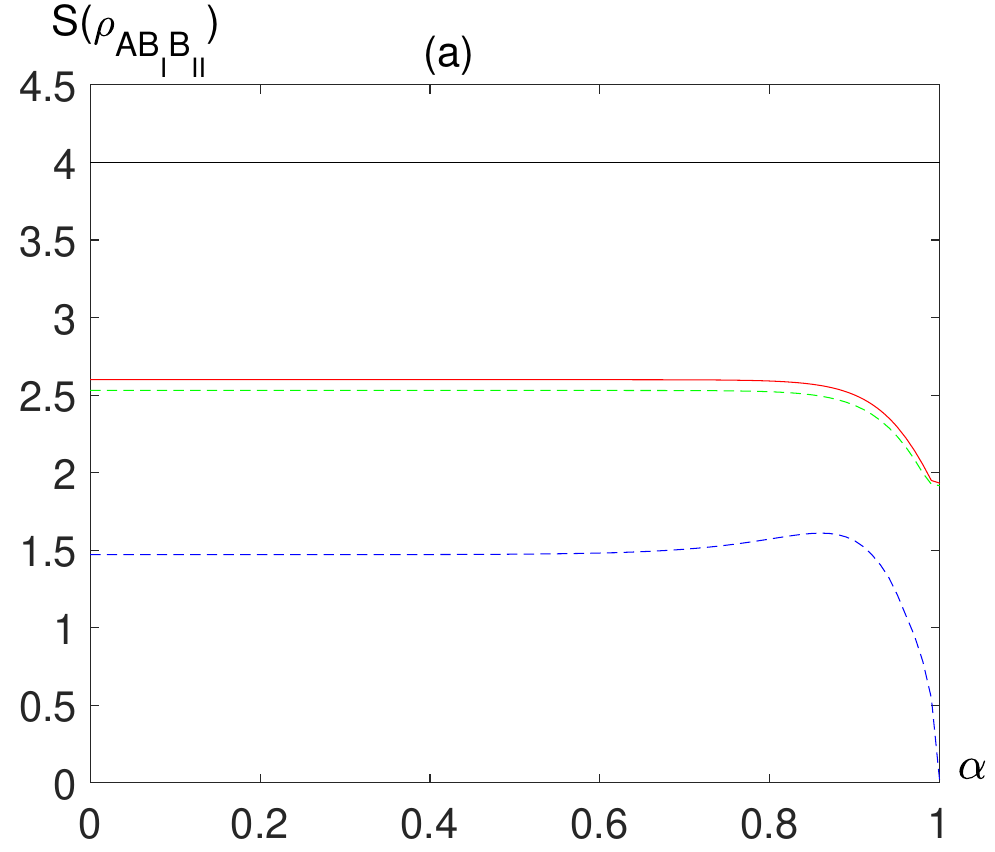}
\includegraphics[width=6cm]{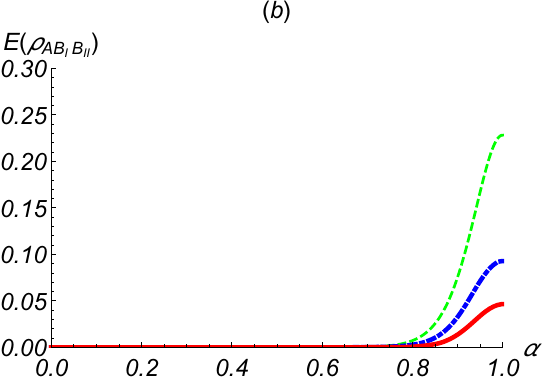}
\caption{\label{fig:fig3} 
 (a) GTN $S(\rho_{AB_IB_{II}})$ of the three-qubit system as a function of the dilaton parameter $\alpha$ with $\omega=1$, $M=1$ for $r=0.05$, $p=0.05$ (red solid line), $r=0.1$, $p=0.1$ (green dashed line) and $r=0.5$, $p=0.5$ (blue dashed line). (b) GTE $E(\rho_{AB_IB_{II}})$ of the three-qubit system as a function of the dilaton parameter $\alpha$ with $\omega=1$, $M=1$ for $r=0.1$, $p=0.5$ (green dashed line), $r=0.5$, $p=0.5$ (blue dashed line) and $r=0.7$, $p=0.5$ (red solid line).}
\end{center}
\end{figure}

In Fig.3, we shows how the dilaton parameter $\alpha$ affects the GTN $S(\rho_{A B_{\uppercase\expandafter{\romannumeral1}} B_{\uppercase\expandafter{\romannumeral2}} })$ and GTE $E(\rho_{A B_{\uppercase\expandafter{\romannumeral1}} B_{\uppercase\expandafter{\romannumeral2}} })$ in the physically inaccessible region for different decoherence strength $r$. It is evident from Fig.3(a) that $S(\rho_{A B_{\uppercase\expandafter{\romannumeral1}} B_{\uppercase\expandafter{\romannumeral2}} })$ always remains below $4$ for every value of $\alpha$, namely, GTN does not arise among modes $A$, $B_{\uppercase\expandafter{\romannumeral1}}$, and $B_{\uppercase\expandafter{\romannumeral2}}$ in the physically inaccessible region. Furthermore, Fig.3(b) shows that the GTE $E(\rho_{A B_{\uppercase\expandafter{\romannumeral1}} B_{\uppercase\expandafter{\romannumeral2}} })$ is initially absent and appears only after $\alpha$ exceeds a relatively large threshold. These results indicate that even when modes $A$, $B_{\uppercase\expandafter{\romannumeral1}}$, and $B_{\uppercase\expandafter{\romannumeral2}}$ are separated by the black hole event horizon, the dilaton effect of the black hole can generate physically inaccessible GTE among them. From a physical mechanism perspective, this process can be interpreted as a transfer of entanglement. Specially, modes $A$, $B_{\uppercase\expandafter{\romannumeral1}}$, and $C_{\uppercase\expandafter{\romannumeral1}}$ share GTE initially, then, as the dilaton effect of the black hole intensifies, it induces an interaction between $B_{\uppercase\expandafter{\romannumeral1}}$ and $B_{\uppercase\expandafter{\romannumeral2}}$ that transfers quantum information from $B_{\uppercase\expandafter{\romannumeral1}}$ to $B_{\uppercase\expandafter{\romannumeral2}}$, and thereby produces GTE among $A$, $B_{\uppercase\expandafter{\romannumeral1}}$, and $B_{\uppercase\expandafter{\romannumeral2}}$. Comparing the three curves in the figure, we find that both the growth amplitude and rate of GTE $E(\rho_{A B_{\uppercase\expandafter{\romannumeral1}} B_{\uppercase\expandafter{\romannumeral2}}})$ decrease as the decoherence strength $r$ increases. This trend implies that a larger fraction of the initial GTE among modes $A$, $B_{\uppercase\expandafter{\romannumeral1}}$, and $C_{\uppercase\expandafter{\romannumeral1}}$  is transferred to the environment, while the GTE generated among modes $A$, $B_{\uppercase\expandafter{\romannumeral1}}$, and $B_{\uppercase\expandafter{\romannumeral2}}$ diminishes. That is to say, environmental decoherence inhibits the generation of GTE among modes $A$, $B_{\uppercase\expandafter{\romannumeral1}}$, and $B_{\uppercase\expandafter{\romannumeral2}}$.

\begin{figure}
\begin{center}
\includegraphics[width=6cm]{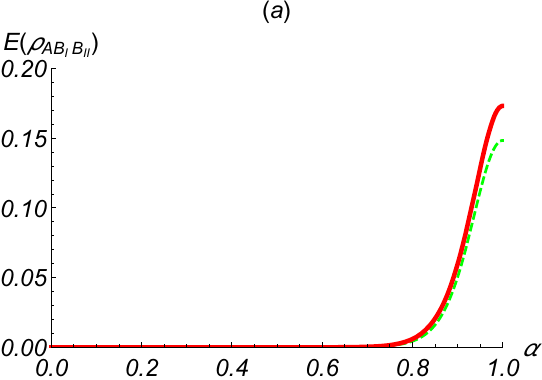}
\includegraphics[width=6cm]{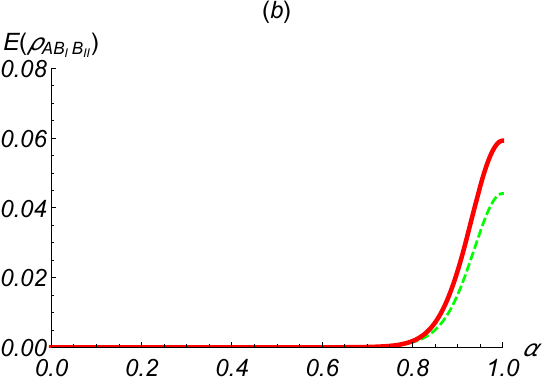}
\caption{\label{fig:fig4} 
 (a) GTE $E(\rho_{AB_IB_{II}})$ of the three-qubit system as a function of the dilaton parameter $\alpha$ with $\omega=1$, $M=1$ for $r=0.3$, $p=0.1$ (green dashed line), and $r=0.3$, $p=0.9$ (red solid line). (b) GTE $E(\rho_{AB_IB_{II}})$ of the three-qubit system as a function of the dilaton parameter $\alpha$ with $\omega=1$, $M=1$ for $r=0.7$, $p=0.1$ (green dashed line), and $r=0.7$, $p=0.9$ (red solid line). }
\end{center}
\end{figure}

In Fig.4, we present the influence of the decoherence parameter $p$ on the GTE $E(\rho_{A B_{\uppercase\expandafter{\romannumeral1}} B_{\uppercase\expandafter{\romannumeral2}}} )$. It is found that the GTE $E(\rho_{A B_{\uppercase\expandafter{\romannumeral1}} B_{\uppercase\expandafter{\romannumeral2}}} )$ increases as $p$ increases. In particular, this increase is pronounced when $p$ is relatively small. This result corresponds closely to the dynamic evolution of the GTE $E(\rho_{A B_{\uppercase\expandafter{\romannumeral1}} C_{\uppercase\expandafter{\romannumeral1}}} )$ shown in Fig.2 Specifically, a larger the decoherence coefficient $p$ leads to more outflow of the initial GTE shared among modes $A$, $B_{\uppercase\expandafter{\romannumeral1}}$, and $C_{\uppercase\expandafter{\romannumeral1}}$. When the decoherence strength $r$ of the environment is fixed, a greater portion of this outflowing GTE is transferred to modes $A$, $B_{\uppercase\expandafter{\romannumeral1}}$, and $B_{\uppercase\expandafter{\romannumeral2}}$, thereby enhancing the GTE $E(\rho_{A B_{\uppercase\expandafter{\romannumeral1}} B_{\uppercase\expandafter{\romannumeral2}}} )$ generated among them.

In the following, we focus on the evolution of GTN and GTE for the modes $A$, $B_{\uppercase\expandafter{\romannumeral2}}$, and $C_{\uppercase\expandafter{\romannumeral2}}$ in the physically inaccessible region. Tracing out the modes $B_{\uppercase\expandafter{\romannumeral1}}$ and $C_{\uppercase\expandafter{\romannumeral1}}$ in the density matrix defined by Eq.(25), we can derive the reduced density matrix $\rho_{A B_{\uppercase\expandafter{\romannumeral2}} C_{\uppercase\expandafter{\romannumeral2}}}$, which is given by 	
\begin{eqnarray}
\rho_{A B_{\uppercase\expandafter{\romannumeral2}} C_{\uppercase\expandafter{\romannumeral2}}}
= \left( \begin{array}{cccccccc}
 \eta_{11}& 0 & 0 &0& 0 & \eta_{16}& \eta_{17} & 0 \\
  0& \eta_{22}& \eta_{23} & 0& 0 & 0& 0 & \eta_{28}\\
  0& \eta_{32} & \eta_{33} & 0& 0 & 0& 0 & \eta_{38}\\
 0 & 0 &0  &\eta_{44}& 0 & 0& 0 & 0\\
 0 & 0 &0  &0& \eta_{55} & 0& 0 & 0\\
\eta_{61}& 0 &0  &0& 0 & \eta_{66}&\eta_{67} & 0\\
\eta_{71} & 0 &0  &0& 0 & \eta_{76}& \eta_{77} & 0\\
 0 & \eta_{82} &\eta_{83} &0& 0 & 0& 0 & \eta_{88}
\end{array} \right),\nonumber\\
\end{eqnarray}
with
\begin{eqnarray*}
&&\eta_{11}=\frac{1}{3}\beta ^{2}(2+((2+\beta ^{2})p-2)r),\nonumber\\
&&\eta_{16}=\eta_{17}=\eta_{61}=\eta_{71}=\frac{1}{3}\beta ^{2}\sqrt{1-\beta ^{2}}\sqrt{1-r},\nonumber\\
&&\eta_{22}=\eta_{33}=\frac{1}{3}(1-\beta ^{2})(1+(p\beta ^{2}+p-1)r),\nonumber\\
&&\eta_{23}=\eta_{32}=\frac{1}{3}(1-\beta ^{2})(1+(p-1)r),\nonumber\\
&&\eta_{28}=\eta_{38}=\eta_{82}=\eta_{83}=\frac{1}{3}(1-\beta ^{2})^{\frac{3}{2}}\sqrt{1-r},\nonumber\\
&&\eta_{44}=\frac{1}{3}(\beta ^{2}-1)^{2}pr,\nonumber\\
&&\eta_{55}=\frac{1}{3}(\beta ^{4}-\beta ^{2}((\beta ^{2}   +2)p-2)r),\nonumber\\
&&\eta_{66}=\eta_{77}=\frac{1}{3}(\beta ^{2}-1)(r(p-1)+ \beta ^{2}(pr-1)),\nonumber\\
&&\eta_{67}=\eta_{76}=\frac{1}{3}(\beta ^{2}-1)(p-1)r,\nonumber\\
&&\eta_{88}=-\frac{1}{3}(\beta ^{2}-1)^{2}(pr-1),\nonumber\\
&&\beta=(e^{-8\pi\omega(M-\alpha)}+1)^{-\frac{1}{2}}.
\end{eqnarray*}

Similarly, we evaluate the GTN and GTE for modes $A$, $B_{\uppercase\expandafter{\romannumeral2}}$, and $C_{\uppercase\expandafter{\romannumeral2}}$ in the physically inaccessible region by employing the methods described in the preceding section.

\begin{figure}
\begin{center}
\includegraphics[width=6cm]{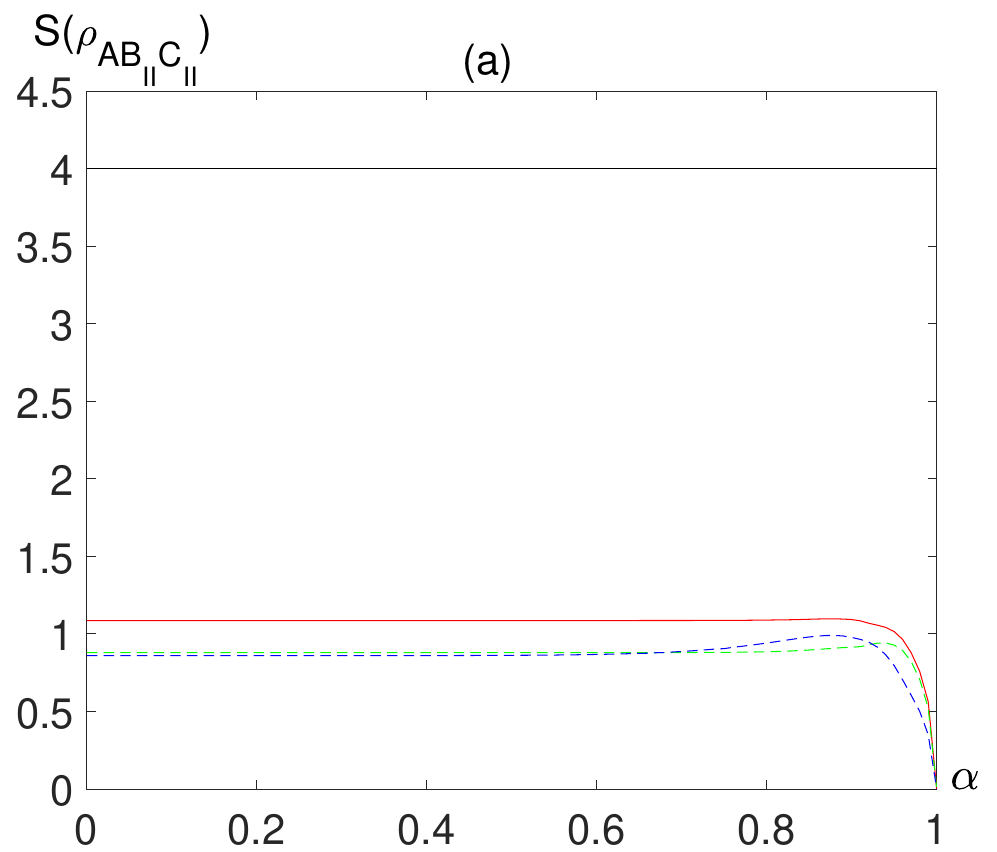}
\includegraphics[width=6cm]{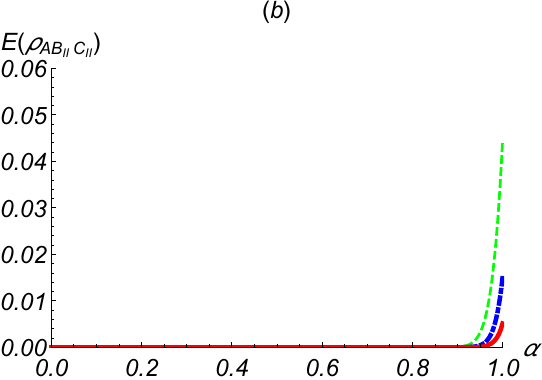}
\caption{\label{fig:fig5} 
 (a) GTN $S(\rho_{AB_{II}C_{II}} )$ of the three-qubit system as a function of the dilaton parameter $\alpha$ with $\omega=1$, $M=1$ for $r=0.05$, $p=0.05$ (red solid line), $r=0.1$, $p=0.1$ (green dashed line) and $r=0.5$, $p=0.5$ (blue dashed line). (b) GTE $E(\rho_{AB_{II}C_{II}} )$ of the three-qubit system as a function of the dilaton parameter $\alpha$ with $\omega=1$, $M=1$ for $r=0.1$, $p=0.5$ (green dashed line), $r=0.4$, $p=0.5$ (blue dashed line) and $r=0.6$, $p=0.5$ (red solid line).}
\end{center}
\end{figure}

In Fig.5, we illustrates the GTN $S(\rho_{A B_{\uppercase\expandafter{\romannumeral2}} C_{\uppercase\expandafter{\romannumeral2}} })$ and the GTE $E(\rho_{A B_{\uppercase\expandafter{\romannumeral2}} C_{\uppercase\expandafter{\romannumeral2}} })$ in the physically inaccessible region as functions of the dilaton parameter $\alpha$ for various decoherence strengths $r$. As shown in Fig.5(a), although the GTN $S(\rho_{A B_{\uppercase\expandafter{\romannumeral2}} C_{\uppercase\expandafter{\romannumeral2}} })$ remains nonzero for any value of $\alpha$, it consistently falls below $4$. This finding indicates that no GTN is generated among the modes $A$, $B_{\uppercase\expandafter{\romannumeral2}}$, and $C_{\uppercase\expandafter{\romannumeral2}}$ in the physically inaccessible region, which means that GTN cannot cross the event horizon of the black hole. Furthermore, it is observed from Fig.5(b) that the dilaton effect of the black hole can generate GTE $E(\rho_{A B_{\uppercase\expandafter{\romannumeral2}} C_{\uppercase\expandafter{\romannumeral2}} })$ in a physically inaccessible region. Notably, the critical value of the dilaton parameter $\alpha$at which this GTE begins to emerge is larger than that for the GTE $E(\rho_{A B_{\uppercase\expandafter{\romannumeral2}} C_{\uppercase\expandafter{\romannumeral2}} })$ shown in Fig.3(b) A comparison of the three curves further reveals that increasing the decoherence strength $r$ not only delays the critical $\alpha$ for GTE $E(\rho_{A B_{\uppercase\expandafter{\romannumeral2}} C_{\uppercase\expandafter{\romannumeral2}} })$, but also suppresses the growth rate of the GTE. This observation suggests that environmental decoherence inhibits the transfer of quantum entanglement from the physically accessible region to the physically inaccessible region, thereby restricting the generation of GTE $E(\rho_{A B_{\uppercase\expandafter{\romannumeral2}} C_{\uppercase\expandafter{\romannumeral2}} })$.

\begin{figure}
\begin{center}
\includegraphics[width=6cm]{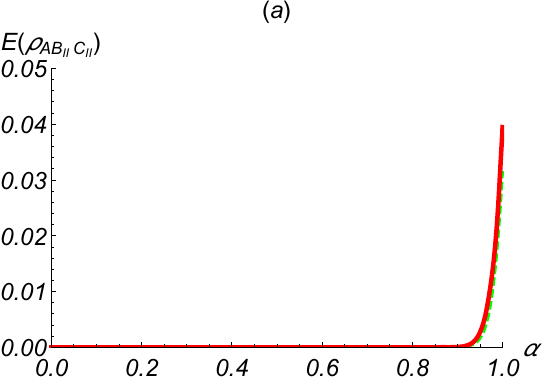}
\includegraphics[width=6cm]{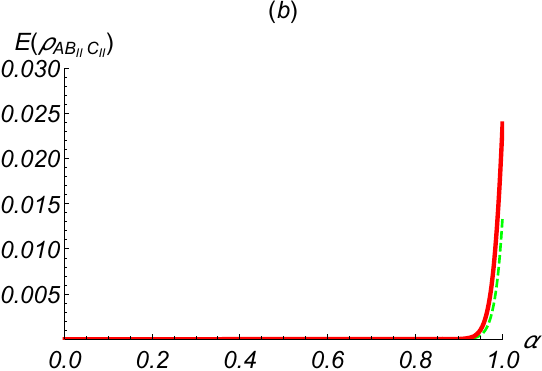}
\caption{\label{fig:fig6} 
 (a) GTE $E(\rho_{AB_{II}C_{II}})$ of the three-qubit system as a function of the dilaton parameter $\alpha$ with $\omega=1$, $M=1$ for $r=0.2$, $p=0.1$ (green dashed line), and $r=0.2$, $p=0.9$ (red solid line). (b) GTE $E(\rho_{AB_{II}C_{II}})$ of the three-qubit system as a function of the dilaton parameter $\alpha$ with $\omega=1$, $M=1$ for $r=0.4$, $p=0.1$ (green dashed line), and $r=0.4$, $p=0.9$ (red solid line). }
\end{center}
\end{figure}

In Fig.6, we plot GTE $E(\rho_{A B_{\uppercase\expandafter{\romannumeral2}} C_{\uppercase\expandafter{\romannumeral2}} })$ as a function of the dilaton parameter $\alpha$ for different decoherence parameter $p$.  It is clear that  regardless of the value of $p$, the GTE $E(\rho_{A B_{\uppercase\expandafter{\romannumeral2}} C_{\uppercase\expandafter{\romannumeral2}} })$ increases only slightly with $p$. This observation indicates that the influence of the decoherence parameter $p$ on GTE $E(\rho_{A B_{\uppercase\expandafter{\romannumeral2}} C_{\uppercase\expandafter{\romannumeral2}} })$ is exceedingly weak, in contrast to its effect on GTE $E(\rho_{A B_{\uppercase\expandafter{\romannumeral1}} B_{\uppercase\expandafter{\romannumeral2}} })$ shown in Fig.4. The underlying physical reason is that only a limited amount of the initial GTE that shared between modes $A$, $B_{\uppercase\expandafter{\romannumeral2}}$, and $C_{\uppercase\expandafter{\romannumeral1}}$ is transferred to the subspace comprising modes $A$, $B_{\uppercase\expandafter{\romannumeral2}}$, and $C_{\uppercase\expandafter{\romannumeral2}}$. Consequently, the generation of GTE $E(\rho_{A B_{\uppercase\expandafter{\romannumeral2}} C_{\uppercase\expandafter{\romannumeral2}} })$ is minimal, resulting in a correspondingly weak dependence on the decoherence parameter $p$.

Finally, by taking the trace of the modes $B_{\uppercase\expandafter{\romannumeral2}}$ and $C_{\uppercase\expandafter{\romannumeral1}}$ of the density matrix in Eq.(25), we can obtain the reduced density matrix $\rho_{A B_{\uppercase\expandafter{\romannumeral1}} C_{\uppercase\expandafter{\romannumeral2}} }$ for the physically inaccessible region, which reads  

\begin{equation}
\rho_{A B_{\uppercase\expandafter{\romannumeral1}} C_{\uppercase\expandafter{\romannumeral2}} }=\left( \begin{array}{cccccccc}
 \varepsilon_{11}
 & 0 & 0 &\varepsilon_{14}
& 0 & \varepsilon_{16}
& 0 & 0 \\
  0& \varepsilon_{22}
& 0 & 0& 0 & 0& 0 & 0\\
  0& 0 & \varepsilon_{33}
 & 0& \varepsilon_{35}
 & 0& 0 & \varepsilon_{38}
\\
\varepsilon_{41}
& 0 &0  & \varepsilon_{44}
& 0 & \varepsilon_{46}
& 0 & 0\\
 0 & 0 &\varepsilon_{53}
  &0& \varepsilon_{55}
 & 0& 0 & \varepsilon_{58}
\\
 \varepsilon_{61}
 & 0 &0  &\varepsilon_{64}
& 0 & \varepsilon_{66}
& 0 & 0\\
 0 &0 &0  &0& 0 & 0& \varepsilon_{77}
 & 0\\
 0 & 0 &\varepsilon_{83}
  &0& \varepsilon_{85}
 & 0& 0 & \varepsilon_{88}
\end{array} \right),
\end{equation}
with
\begin{eqnarray*}
&&\varepsilon_{11}=\frac{1}{3}\beta ^{2}(1+(\beta ^{2}p+p-1)r),\nonumber\\
&&\varepsilon_{14}=\varepsilon_{41}=\frac{1}{3}\beta \sqrt{1-\beta ^{2}}(1+(p-1)r),\nonumber\\
&&\varepsilon_{16}=\varepsilon_{61}=\frac{1}{3}\beta^{2} \sqrt{1-\beta ^{2}}\sqrt{1-r},\nonumber\\
&&\varepsilon_{22}=\frac{1}{3}\beta ^{2}(1-\beta ^{2})pr,\nonumber\\
&&\varepsilon_{33}=\frac{1}{3}(1+((1+\beta ^{2}-\beta ^{4})p-1)r),\nonumber\\
&&\varepsilon_{35}=\varepsilon_{53}=\frac{1}{3}\beta^{3}\sqrt{1-r},\nonumber\\
&&\varepsilon_{38}=\varepsilon_{83}=\frac{1}{3}(1-\beta^{2})^{\frac{3}{2}}\sqrt{1-r},\nonumber\\
\end{eqnarray*}
\begin{eqnarray*}
&&\varepsilon_{44}=\frac{1}{3}(\beta ^{2}-1)(r-1+(\beta ^{2}-2)pr),\nonumber\\
&&\varepsilon_{46}=\varepsilon_{64}=\frac{1}{3}(1-\beta ^{2})\sqrt{1-r},\nonumber\\
&&\varepsilon_{55}=\frac{1}{3}(\beta ^{4}-\beta ^{2}r(\beta ^{2}p+p-1    )),\nonumber\\
&&\varepsilon_{58}=\varepsilon_{85}=\frac{1}{3}(1-p)r \beta \sqrt{1-\beta ^{2}},\nonumber\\
&&\varepsilon_{66}=\frac{1}{3} \beta ^{2}(\beta ^{2}-1)(pr-1),\nonumber\\
&&\varepsilon_{77}=\frac{1}{3}(r-pr+\beta ^{2}(\beta ^{2}-1)(pr-1)),\nonumber\\
&&\varepsilon_{88}=\frac{1}{3}(1-\beta ^{2})(1+r-2pr+\beta ^{2}(pr-1)),\nonumber\\
&&\beta=(e^{-8\pi\omega(M-\alpha)}+1)^{-\frac{1}{2}}.
\end{eqnarray*}

Employing the same method as before, we derive the GTN $S(\rho_{A B_{\uppercase\expandafter{\romannumeral1}} C_{\uppercase\expandafter{\romannumeral2}} })$ and GTE $E(\rho_{A B_{\uppercase\expandafter{\romannumeral1}} C_{\uppercase\expandafter{\romannumeral2}} })$ associated with the density matrix $\rho_{A B_{\uppercase\expandafter{\romannumeral1}} C_{\uppercase\expandafter{\romannumeral2}} }$. In addition, by virtue of the exchange symmetry between Bob and Charlie, it follows that the GTN and GTE for  the density matrix $\rho_{A B_{\uppercase\expandafter{\romannumeral2}} C_{\uppercase\expandafter{\romannumeral1}} }$ are identical to those for $\rho_{A B_{\uppercase\expandafter{\romannumeral1}} C_{\uppercase\expandafter{\romannumeral2}} }$. Hence, we focus our discussion solely on the GTN and GTE of the density matrix $\rho_{A B_{\uppercase\expandafter{\romannumeral1}} C_{\uppercase\expandafter{\romannumeral2}} }$.

\begin{figure}
\begin{center}
\includegraphics[width=6cm]{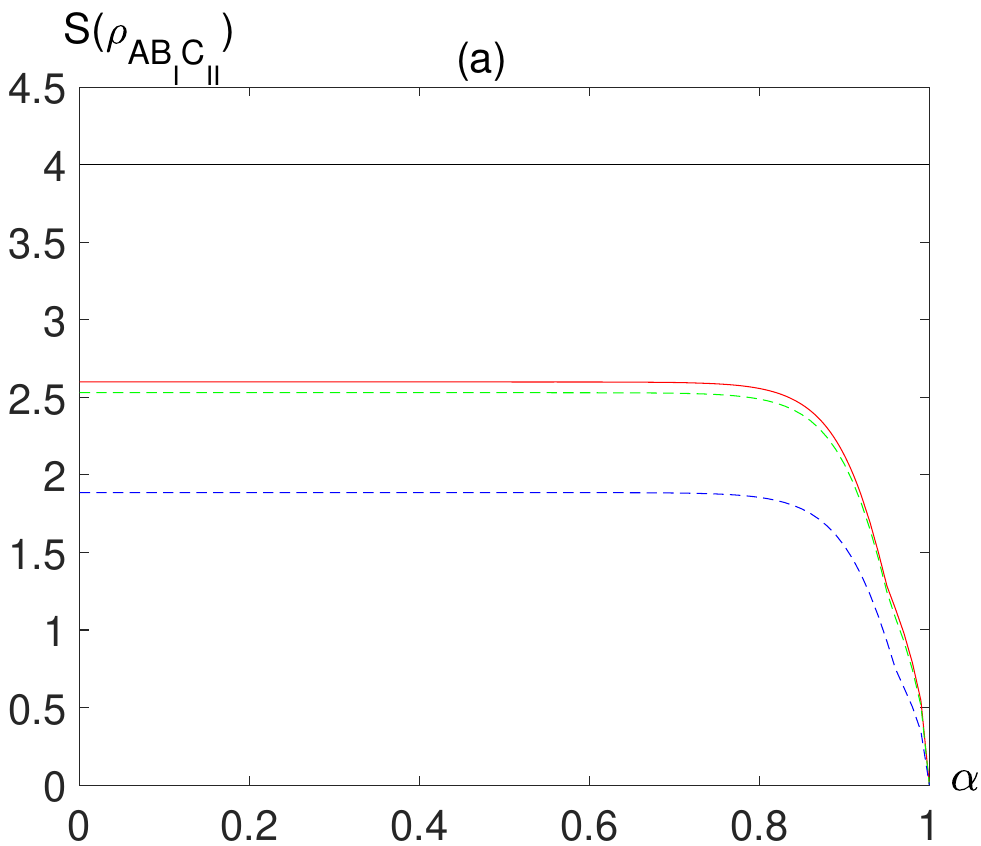}
\includegraphics[width=6cm]{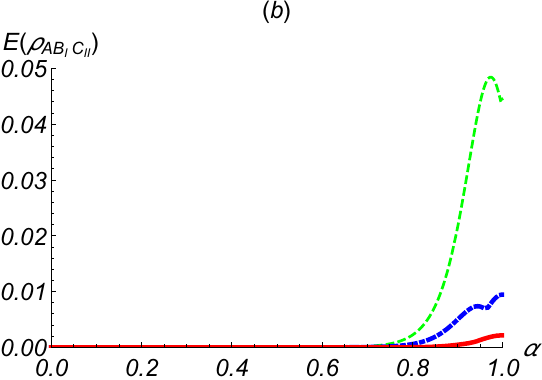}
\caption{\label{fig:fig7} 
 (a) GTN $S(\rho_{AB_IC_{II}} )$ of the three-qubit system as a function of the dilaton parameter $\alpha$ with $\omega=1$, $M=1$ for $r=0.05$, $p=0.05$ (red solid line), $r=0.1$, $p=0.1$ (green dashed line) and $r=0.5$, $p=0.5$ (blue dashed line). (b) GTE $E(\rho_{AB_IC_{II}} )$ of the three-qubit system as a function of the dilaton parameter $\alpha$ with $\omega=1$, $M=1$ for $r=0.1$, $p=0.5$ (green dashed line), $r=0.5$, $p=0.5$ (blue dashed line) and $r=0.7$, $p=0.5$ (red solid line). }
\end{center}
\end{figure}

In Fig.7, we plot the GTN $S(\rho_{A B_{\uppercase\expandafter{\romannumeral1}} C_{\uppercase\expandafter{\romannumeral2}} })$ and GTE $E(\rho_{A B_{\uppercase\expandafter{\romannumeral1}} C_{\uppercase\expandafter{\romannumeral2}} })$ as functions of the dilaton parameter $\alpha$ for various decoherence strengths $r$. The results reveal that the dilaton effect of the black hole can still generate GTE among the physically inaccessible modes $A$, $B_{\uppercase\expandafter{\romannumeral1}}$, and $C_{\uppercase\expandafter{\romannumeral2}}$, while GTN cannot be established. The underlying physical mechanism for the generation of $E(\rho_{A B_{\uppercase\expandafter{\romannumeral1}} C_{\uppercase\expandafter{\romannumeral2}} })$ is the transfer of initial entanglement. Moreover, the analysis of the three curves in Fig.7(b) indicates that as the decoherence strength $r$ increases, both the growth rate and the magnitude of the GTE $E(\rho_{A B_{\uppercase\expandafter{\romannumeral1}} C_{\uppercase\expandafter{\romannumeral2}} })$ diminish, suggesting that environmental decoherence suppresses the generation of GTE among modes $A$, $B_{\uppercase\expandafter{\romannumeral1}}$, and $C_{\uppercase\expandafter{\romannumeral2}}$.

\begin{figure}
\begin{center}
\includegraphics[width=6cm]{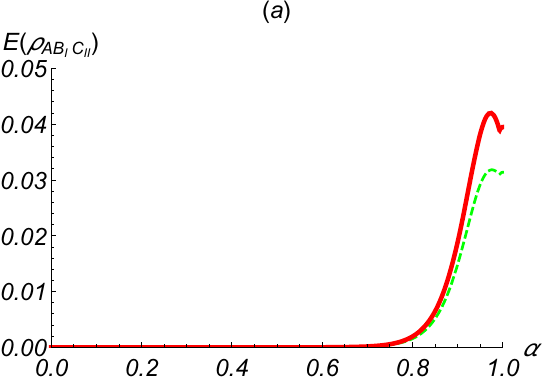}
\includegraphics[width=6cm]{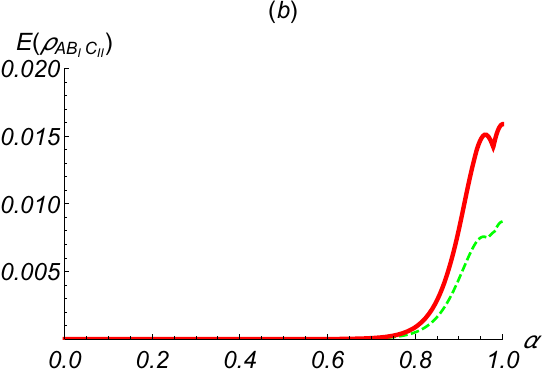}
\caption{\label{fig:fig8} 
 (a)  GTE $E(\rho_{AB_IC_{II}})$ of the three-qubit system as a function of the dilaton parameter $\alpha$ with $\omega=1$, $M=1$ for $r=0.2$, $p=0.1$ (green dashed line), and $r=0.2$, $p=0.9$ (red solid line).
 (b) GTE $E(\rho_{AB_IC_{II}})$ of the three-qubit system as a function of the dilaton parameter $\alpha$ with  $\omega=1$, $M=1$ for $r=0.5$, $p=0.1$ (green dashed line), and $r=0.5$, $p=0.9$ (red solid line).}
\end{center}
\end{figure}

In Figure 8, we plot the GTE $E(\rho_{A B_{\uppercase\expandafter{\romannumeral1}} C_{\uppercase\expandafter{\romannumeral2}} })$ as a function of the dilaton parameter $\alpha$ for different decoherence parameters $p$. From the figure, we find that both the magnitude and the growth rate of $E(\rho_{A B_{\uppercase\expandafter{\romannumeral1}} C_{\uppercase\expandafter{\romannumeral2}} })$ increase with the decoherence parameter $p$, although the rate of increase diminishes for the large values of the decoherence strength $r$. A comparison with Fig. 4 reveals that $E(\rho_{A B_{\uppercase\expandafter{\romannumeral1}} B_{\uppercase\expandafter{\romannumeral2}} })$ is more sensitive to variations in $p$ than  $E(\rho_{A B_{\uppercase\expandafter{\romannumeral1}} C_{\uppercase\expandafter{\romannumeral2}} })$. This observation implies that, under decoherence, a larger fraction of the initial GTE among modes $A$, $B_{\uppercase\expandafter{\romannumeral1}}$, and $C_{\uppercase\expandafter{\romannumeral1}}$ is transferred to the subsystem composed of modes $A$, $B_{\uppercase\expandafter{\romannumeral1}}$, and $B_{\uppercase\expandafter{\romannumeral2}}$.

\section {Bell nonlocality and entanglement for bipartite systems}
Our previous analysis demonstrated that the GTN in the physically accessible region eventually vanishes completely as the dilaton parameter $\alpha$ increases. Furthermore, GTN cannot arise in the remaining physically inaccessible regions. Therefore, to determine whether the loss of GTN in the physically accessible region is redistributed as bipartite nonlocality among subsystems, we next examine how the dilaton effect of the black hole and environmental decoherence affect bipartite quantum nonlocality. Additionally, to clarify the flow of quantum entanglement in the system, we further analyze the dynamic behavior of entanglement among the different bipartite subsystems. We quantify bipartite nonlocality and entanglement using the maximal Bell-CHSH violation and concurrence, respectively. By tracing out any three modes in the density matrix of Eq.(25), we derive the reduced density operators for every bipartite subsystem, with the explicit forms listed as follows
\begin{eqnarray}
\rho_{A B_{\uppercase\expandafter{\romannumeral1}}}=\rho_{A C_{\uppercase\expandafter{\romannumeral1}}}
=\left( \begin{array}{cccc}
\rho^{A B_{\uppercase\expandafter{\romannumeral1}}}_{11} & 0 & 0 &0 \\
  0& \rho^{A B_{\uppercase\expandafter{\romannumeral1}}}_{22} & \rho^{A B_{\uppercase\expandafter{\romannumeral1}}}_{23} & 0\\
  0& \rho^{A B_{\uppercase\expandafter{\romannumeral1}}}_{32} & \rho^{A B_{\uppercase\expandafter{\romannumeral1}}}_{33} & 0\\
0 & 0 &0  &\rho^{A B_{\uppercase\expandafter{\romannumeral1}}}_{44}
\end{array} \right),
\end{eqnarray}
with
\begin{eqnarray*}
&&\rho^{A B_{\uppercase\expandafter{\romannumeral1}}}_{11}=\frac{1}{3}\beta^{2}(1+(2p-1)r),\nonumber\\
&&\rho^{A B_{\uppercase\expandafter{\romannumeral1}}}_{22}=\frac{1}{3}(2+(3p-2)r+\beta^{2}(r-2pr-1)),\nonumber\\
&&\rho^{A B_{\uppercase\expandafter{\romannumeral1}}}_{23}=\rho^{A B_{\uppercase\expandafter{\romannumeral1}}}_{32}= \frac{1}{3} \beta \sqrt{1-r},\nonumber\\
&&\rho^{A B_{\uppercase\expandafter{\romannumeral1}}}_{33}=\frac{1}{3} \beta^{2}(r-2pr+1),\nonumber\\
&&\rho^{A B_{\uppercase\expandafter{\romannumeral1}}}_{44}=\frac{1}{3}(1+(2-3p)r+\beta^{2}(
 (2p-1)r)-1),\nonumber\\
&&\beta=(e^{-8\pi\omega(M-\alpha)}+1)^{-\frac{1}{2}}.
\end{eqnarray*}
\begin{eqnarray}
\rho_{A B_{\uppercase\expandafter{\romannumeral2}}}=\rho_{A C_{\uppercase\expandafter{\romannumeral2}}}
=\left( \begin{array}{cccc}
\rho^{A B_{\uppercase\expandafter{\romannumeral2}}}_{11} & 0 & 0 &\rho^{A B_{\uppercase\expandafter{\romannumeral2}}}_{14} \\
  0&\rho^{A B_{\uppercase\expandafter{\romannumeral2}}}_{22} & 0 & 0\\
  0& 0 & \rho^{A B_{\uppercase\expandafter{\romannumeral2}}}_{33} & 0\\
\rho^{A B_{\uppercase\expandafter{\romannumeral2}}}_{41}& 0 &0  &\rho^{A B_{\uppercase\expandafter{\romannumeral2}}}_{44}
\end{array} \right),
\end{eqnarray}
with
\begin{eqnarray*}
&&\rho^{A B_{\uppercase\expandafter{\romannumeral1}}}_{11}=\frac{1}{3}(1+\beta^{2}+(p-1+\beta^{2}(2p-1))r),\nonumber\\
&&\rho^{A B_{\uppercase\expandafter{\romannumeral1}}}_{14}=\rho^{A B_{\uppercase\expandafter{\romannumeral1}}}_{41}=\frac{1}{3}\sqrt{1-\beta^{2}}\sqrt{1-r},\nonumber\\
&&\rho^{A B_{\uppercase\expandafter{\romannumeral1}}}_{22}=\frac{1}{3}(\beta^{2}-1)(1+(2p-1)r),\nonumber\\
&&\rho^{A B_{\uppercase\expandafter{\romannumeral1}}}_{33}=\frac{1}{3}(r-pr+\beta^{2}(1+r-2pr)),\nonumber\\
&&\rho^{A B_{\uppercase\expandafter{\romannumeral1}}}_{44}=\frac{1}{3}(\beta^{2}-1)((2p-1)r-1),\nonumber\\
&&\beta=(e^{-8\pi\omega(M-\alpha)}+1)^{-\frac{1}{2}}.
\end{eqnarray*}

\begin{eqnarray}
\rho_{B_{\uppercase\expandafter{\romannumeral1}}  C_{\uppercase\expandafter{\romannumeral2}}}=\left( \begin{array}{cccc}
\rho^{B_{\uppercase\expandafter{\romannumeral1}}  C_{\uppercase\expandafter{\romannumeral2}}}_{11} & 0 & 0 &\rho^{B_{\uppercase\expandafter{\romannumeral1}}  C_{\uppercase\expandafter{\romannumeral2}}}_{14} \\
  0&\rho^{B_{\uppercase\expandafter{\romannumeral1}}  C_{\uppercase\expandafter{\romannumeral2}}}_{22} & 0 & 0\\
  0& 0 & \rho^{B_{\uppercase\expandafter{\romannumeral1}}  C_{\uppercase\expandafter{\romannumeral2}}}_{33} & 0\\
 \rho^{B_{\uppercase\expandafter{\romannumeral1}}  C_{\uppercase\expandafter{\romannumeral2}}}_{41} & 0 &0  &\rho^{B_{\uppercase\expandafter{\romannumeral1}}  C_{\uppercase\expandafter{\romannumeral2}}}_{44}
\end{array} \right),
\end{eqnarray}
with
\begin{eqnarray*}
&&\rho^{B_{\uppercase\expandafter{\romannumeral1}}  C_{\uppercase\expandafter{\romannumeral2}}}_{11}=\frac{1}{3}(\beta^{2}+\beta^{4}),\nonumber\\
&&\rho^{B_{\uppercase\expandafter{\romannumeral1}}  C_{\uppercase\expandafter{\romannumeral2}}}_{14}=\rho^{B_{\uppercase\expandafter{\romannumeral1}}  C_{\uppercase\expandafter{\romannumeral2}}}_{41}=\frac{1}{3} \beta \sqrt{1-\beta^{2}},\nonumber\\
&&\rho^{B_{\uppercase\expandafter{\romannumeral1}}  C_{\uppercase\expandafter{\romannumeral2}}}_{22}=\frac{1}{3}(\beta^{2}-\beta^{4}),\nonumber\\
&&\rho^{B_{\uppercase\expandafter{\romannumeral1}}  C_{\uppercase\expandafter{\romannumeral2}}}_{33}=\frac{1}{3}(1+\beta^{2}-\beta^{4}),\nonumber\\
&&\rho^{B_{\uppercase\expandafter{\romannumeral1}}  C_{\uppercase\expandafter{\romannumeral2}}}_{44}=\frac{1}{3}(2-3\beta^{2}+\beta^{4}),\nonumber\\
&&\beta=(e^{-8\pi\omega(M-\alpha)}+1)^{-\frac{1}{2}}.
\end{eqnarray*}
\begin{eqnarray}
\rho_{B_{\uppercase\expandafter{\romannumeral1}} C_{\uppercase\expandafter{\romannumeral1}}}=\left( \begin{array}{cccc}
\rho^{B_{\uppercase\expandafter{\romannumeral1}} C_{\uppercase\expandafter{\romannumeral1}}}_{11} & 0 & 0 &0 \\
  0&\rho^{B_{\uppercase\expandafter{\romannumeral1}} C_{\uppercase\expandafter{\romannumeral1}}}_{22} & \rho^{B_{\uppercase\expandafter{\romannumeral1}} C_{\uppercase\expandafter{\romannumeral1}}}_{23} & 0\\
  0& \rho^{B_{\uppercase\expandafter{\romannumeral1}} C_{\uppercase\expandafter{\romannumeral1}}}_{32} & \rho^{B_{\uppercase\expandafter{\romannumeral1}} C_{\uppercase\expandafter{\romannumeral1}}}_{33} & 0\\
0 & 0 &0  &\rho^{B_{\uppercase\expandafter{\romannumeral1}} C_{\uppercase\expandafter{\romannumeral1}}}_{44}
\end{array} \right),
\end{eqnarray}
with
\begin{eqnarray*}
&&\rho^{B_{\uppercase\expandafter{\romannumeral1}} C_{\uppercase\expandafter{\romannumeral1}}}_{11}=\frac{1}{3}\beta^{4},\nonumber\\
&&\rho^{B_{\uppercase\expandafter{\romannumeral1}} C_{\uppercase\expandafter{\romannumeral1}}}_{22}=\rho^{B_{\uppercase\expandafter{\romannumeral1}} C_{\uppercase\expandafter{\romannumeral1}}}_{33}=\frac{1}{3}\beta^{2}(2-\beta^{2}),\nonumber\\
&&\rho^{B_{\uppercase\expandafter{\romannumeral1}} C_{\uppercase\expandafter{\romannumeral1}}}_{23}=\rho^{B_{\uppercase\expandafter{\romannumeral1}} C_{\uppercase\expandafter{\romannumeral1}}}_{32}=\frac{1}{3}\beta^{2},\nonumber\\
&&\rho^{B_{\uppercase\expandafter{\romannumeral1}} C_{\uppercase\expandafter{\romannumeral1}}}_{44}=\frac{1}{3}(\beta^{4}-4\beta^{2}+3),\nonumber\\
&&\beta=(e^{-8\pi\omega(M-\alpha)}+1)^{-\frac{1}{2}}.
\end{eqnarray*}

\begin{eqnarray}
\rho_{B_{\uppercase\expandafter{\romannumeral2}}  C_{\uppercase\expandafter{\romannumeral1}}}=\left( \begin{array}{cccc}
\rho^{B_{\uppercase\expandafter{\romannumeral2}}  C_{\uppercase\expandafter{\romannumeral1}}}_{11}& 0 & 0 &\rho^{B_{\uppercase\expandafter{\romannumeral2}}  C_{\uppercase\expandafter{\romannumeral1}}}_{14} \\
  0& \rho^{B_{\uppercase\expandafter{\romannumeral2}}  C_{\uppercase\expandafter{\romannumeral1}}}_{22} & 0 & 0\\
  0& 0 & \rho^{B_{\uppercase\expandafter{\romannumeral2}}  C_{\uppercase\expandafter{\romannumeral1}}}_{33} & 0\\
\rho^{B_{\uppercase\expandafter{\romannumeral2}}  C_{\uppercase\expandafter{\romannumeral1}}}_{41} & 0 &0  &\rho^{B_{\uppercase\expandafter{\romannumeral2}}  C_{\uppercase\expandafter{\romannumeral1}}}_{44}
\end{array} \right),
\end{eqnarray}
with
\begin{eqnarray*}
&&\rho^{B_{\uppercase\expandafter{\romannumeral2}}  C_{\uppercase\expandafter{\romannumeral1}}}_{11}=\frac{1}{3}(\beta^{2}+\beta^{4}),\nonumber\\
&&\rho^{B_{\uppercase\expandafter{\romannumeral2}}  C_{\uppercase\expandafter{\romannumeral1}}}_{14}=\rho^{B_{\uppercase\expandafter{\romannumeral2}}  C_{\uppercase\expandafter{\romannumeral1}}}_{41}=\frac{1}{3} \beta \sqrt{1-\beta^{2}},\nonumber\\
&&\rho^{B_{\uppercase\expandafter{\romannumeral2}}  C_{\uppercase\expandafter{\romannumeral1}}}_{22}=\frac{1}{3}(1+\beta^{2}-\beta^{4}),\nonumber\\
&&\rho^{B_{\uppercase\expandafter{\romannumeral2}}  C_{\uppercase\expandafter{\romannumeral1}}}_{33}=\frac{1}{3}(\beta^{2}-\beta^{4}),\nonumber\\
&&\rho^{B_{\uppercase\expandafter{\romannumeral2}}  C_{\uppercase\expandafter{\romannumeral1}}}_{44}=\frac{1}{3}(2-3\beta^{2}+\beta^{4}),\nonumber\\
&&\beta=(e^{-8\pi\omega(M-\alpha)}+1)^{-\frac{1}{2}}.
\end{eqnarray*}
\begin{eqnarray}
\rho_{B_{\uppercase\expandafter{\romannumeral1}} B_{\uppercase\expandafter{\romannumeral2}}}=
\rho_{C_{\uppercase\expandafter{\romannumeral1}}  C_{\uppercase\expandafter{\romannumeral2}}}=\left(\begin{array}{cccc}
\rho^{B_{\uppercase\expandafter{\romannumeral1}} B_{\uppercase\expandafter{\romannumeral2}}}_{11} & 0 & 0 &\rho^{B_{\uppercase\expandafter{\romannumeral1}} B_{\uppercase\expandafter{\romannumeral2}}}_{14} \\
  0& 0 & 0 & 0\\
  0& 0 & \rho^{B_{\uppercase\expandafter{\romannumeral1}} B_{\uppercase\expandafter{\romannumeral2}}}_{33} & 0\\
\rho^{B_{\uppercase\expandafter{\romannumeral1}} B_{\uppercase\expandafter{\romannumeral2}}}_{41} & 0 &0  &\rho^{B_{\uppercase\expandafter{\romannumeral1}} B_{\uppercase\expandafter{\romannumeral2}}}_{44}
\end{array} \right),
\end{eqnarray}
with
\begin{eqnarray*}
&&\rho^{B_{\uppercase\expandafter{\romannumeral1}} B_{\uppercase\expandafter{\romannumeral2}}}_{11}=\frac{2}{3} \beta^{2},\nonumber\\
&&\rho^{B_{\uppercase\expandafter{\romannumeral1}} B_{\uppercase\expandafter{\romannumeral2}}}_{14}=\rho^{B_{\uppercase\expandafter{\romannumeral1}} B_{\uppercase\expandafter{\romannumeral2}}}_{41}=\frac{2}{3} \beta \sqrt{1-\beta^{2}},\nonumber\\
&&\rho^{B_{\uppercase\expandafter{\romannumeral1}} B_{\uppercase\expandafter{\romannumeral2}}}_{33}=\frac{1}{3},\nonumber\\
&&\rho^{B_{\uppercase\expandafter{\romannumeral1}} B_{\uppercase\expandafter{\romannumeral2}}}_{44}=\frac{2}{3} (1-\beta^{2}),\nonumber\\
&&\beta=(e^{-8\pi\omega(M-\alpha)}+1)^{-\frac{1}{2}}.
\end{eqnarray*}

\begin{eqnarray}
\rho_{B_{\uppercase\expandafter{\romannumeral2}}  C_{\uppercase\expandafter{\romannumeral2}}}=\left( \begin{array}{cccc}
\rho^{B_{\uppercase\expandafter{\romannumeral2}}  C_{\uppercase\expandafter{\romannumeral2}}}_{11} & 0 & 0 &0 \\
  0& \rho^{B_{\uppercase\expandafter{\romannumeral2}}  C_{\uppercase\expandafter{\romannumeral2}}}_{22} & \rho^{B_{\uppercase\expandafter{\romannumeral2}}  C_{\uppercase\expandafter{\romannumeral2}}}_{23} & 0\\
  0& \rho^{B_{\uppercase\expandafter{\romannumeral2}}  C_{\uppercase\expandafter{\romannumeral2}}}_{32} & \rho^{B_{\uppercase\expandafter{\romannumeral2}}  C_{\uppercase\expandafter{\romannumeral2}}}_{33} & 0\\
0& 0 &0  &\rho^{B_{\uppercase\expandafter{\romannumeral2}}  C_{\uppercase\expandafter{\romannumeral2}}}_{44}
\end{array} \right),
\end{eqnarray}
with
\begin{eqnarray*}
&&\rho^{B_{\uppercase\expandafter{\romannumeral2}}  C_{\uppercase\expandafter{\romannumeral2}}}_{11}=\frac{1}{3} \beta^{2}(2+\beta^{2}),\nonumber\\
&&\rho^{B_{\uppercase\expandafter{\romannumeral2}}  C_{\uppercase\expandafter{\romannumeral2}}}_{22}=\rho^{B_{\uppercase\expandafter{\romannumeral2}}  C_{\uppercase\expandafter{\romannumeral2}}}_{33}=\frac{1}{3} (1-\beta^{4}),\nonumber\\
&&\rho^{B_{\uppercase\expandafter{\romannumeral2}}  C_{\uppercase\expandafter{\romannumeral2}}}_{23}=\rho^{B_{\uppercase\expandafter{\romannumeral2}}  C_{\uppercase\expandafter{\romannumeral2}}}_{32}=\frac{1}{3} (1-\beta^{2}),\nonumber\\
&&\rho^{B_{\uppercase\expandafter{\romannumeral2}}  C_{\uppercase\expandafter{\romannumeral2}}}_{44}=\frac{1}{3} (\beta^{2}-1)^{2},\nonumber\\
&&\beta=(e^{-8\pi\omega(M-\alpha)}+1)^{-\frac{1}{2}}.
\end{eqnarray*}

Substituting these density matrices into Eqs.(3) and (9) respectively, we derive the Bell nonlocality and entanglement concurrence for each of the corresponding bipartite subsystems, with the explicit forms given by

\begin{eqnarray*}
&&BN(\rho_{B_{\uppercase\expandafter{\romannumeral1}}B_{\uppercase\expandafter{\romannumeral2}}})
=BN(\rho_{C_{\uppercase\expandafter{\romannumeral1}}  C_{\uppercase\expandafter{\romannumeral2}}})\nonumber\\
&&=\max\{\frac{8\sqrt{2}}{3} \left |\beta \sqrt{1-\beta^{2}}\right |,\nonumber\\
&&\frac{2}{3}(16|\beta ^{2}-\beta ^{4}|+(2|\beta ^{2}-1|-1+2 \beta \beta ^{\ast })^{2})^{\frac{1}{2}}\},\nonumber\\
&&BN(\rho_{A B_{\uppercase\expandafter{\romannumeral1}}})=BN(\rho_{A C_{\uppercase\expandafter{\romannumeral1}}})\nonumber\\
&&=\max\{\frac{4\sqrt{2}}{3}|\beta\sqrt{1-r}|,\frac{2}{3}(4|\beta^{2}(r-1)|\nonumber\\
&&+(|\beta^{2}(r-2pr+1)|-|\beta|^{2}|2pr-r+1|\nonumber\\
&&+|2+(3p-2)r+\beta ^{2}(r-2pr-1)|\nonumber\\
&&-|1+(2-3p)r+\beta^{2}(2pr-1-r)|)^{2})^{\frac{1}{2}}\},\nonumber\\
&&BN(\rho_{B_{\uppercase\expandafter{\romannumeral1}}C_{\uppercase\expandafter{\romannumeral1}}})\nonumber\\
&&=\max\{\frac{4\sqrt{2}}{3}\beta\beta^{\ast},
\frac{2}{3}(4|\beta|^{4}+(|\beta|^{4}\nonumber\\
&&-2|\beta^{2}(\beta^{2}-2)|+|3-4\beta^{2}+\beta^{4}|)^{2})^{\frac{1}{2}}\},\nonumber\\
\end{eqnarray*}
\begin{eqnarray}
&&BN(\rho_{A B_{\uppercase\expandafter{\romannumeral2}}})=BN(\rho_{A C_{\uppercase\expandafter{\romannumeral2}}})\nonumber\\
&&=\max\{\frac{4\sqrt{2}}{3}|\sqrt{1-\beta^{2}}\sqrt{1-r}|,
\frac{2}{3}(
4|(\beta^{2}-1)(r-1)|\nonumber\\
&&+(|(\beta^{2}-1)(r+1-2pr)|-|(\beta^{2}-1)(1-r+2pr)|\nonumber\\
&&+|\beta^{2}+1+(p-1+\beta^{2}(2p-1))r|-|r-pr\nonumber\\
&&+\beta^{2}(1+r-2pr)|)^{2}
)^{\frac{1}{2}}\},\nonumber\\
&&BN(\rho_{B_{\uppercase\expandafter{\romannumeral1}} C_{\uppercase\expandafter{\romannumeral2}}})=BN(\rho_{B_{\uppercase\expandafter{\romannumeral2}}  C_{\uppercase\expandafter{\romannumeral1}}})\nonumber\\
&&=\max\{\frac{4\sqrt{2}}{3}|\beta\sqrt{1-\beta^{2}}|,
\frac{2}{3}(4|\beta^{2}-\beta^{4}|+
(|\beta^{2}-\beta^{4}|\nonumber\\
&&+|\beta^{2}-\beta^{4}+1|-|2-3\beta^{2}+\beta^{4}|-|\beta^{2}+\beta^{4}|)^{2})^{\frac{1}{2}}\},\nonumber\\
&&BN(\rho_{B_{\uppercase\expandafter{\romannumeral2}} C_{\uppercase\expandafter{\romannumeral2}}})\nonumber\\
&&=\max\{\frac{4\sqrt{2}}{3}|1-\beta^{2}|,
\frac{2}{3}(4|1-\beta^{2}|^{2}+(|\beta^{2}-1|^{2}\nonumber\\
&&+|\beta^{2}(2+\beta^{2})|
-2|1-\beta^{4}|)^{2})^{\frac{1}{2}}\},\nonumber\\
&&\beta=(e^{-8\pi\omega(M-\alpha)}+1)^{-\frac{1}{2}}.
\end{eqnarray}

\begin{eqnarray}
&&C(\rho_{B_{\uppercase\expandafter{\romannumeral1}}  B_{\uppercase\expandafter{\romannumeral2}}})=C(\rho_{C_{\uppercase\expandafter{\romannumeral1}}  C_{\uppercase\expandafter{\romannumeral2}}})\nonumber\\
&&=2\max\{0,-\frac{2}{3}(\beta ^{2}-\beta ^{4})^{\frac{1}{2}}, -\frac{2}{3} \left |\beta (1-\beta^{2})^{\frac{1}{2}}\right | \},\nonumber\\
&&C(\rho_{AB_{\uppercase\expandafter{\romannumeral1}}})=C(\rho_{AC_{\uppercase\expandafter{\romannumeral1}}})\nonumber\\
&&=2\max\{0,-\frac{1}{3}(\beta ^{2}(r-2pr+1)(2+(3p-2)r\nonumber\\
&&+\beta ^{2}(r-1-2pr)))^{\frac{1}{2}},
\frac{1}{3}(-(\beta ^{2}(1+(2p-1)r)
(1+2r\nonumber\\
&&-3pr+\beta ^{2}(2pr-1-r)))^{\frac{1}{2}}
+|\beta(1-r)^{\frac{1}{2}}|)\},\nonumber\\
&&C(\rho_{B_{\uppercase\expandafter{\romannumeral1}}  C_{\uppercase\expandafter{\romannumeral1}}})\nonumber\\
&&=2\max\{0,-\frac{1}{3}(\beta ^{4}(\beta ^{2}-2)^{2})^{\frac{1}{2}},\nonumber\\
&&\frac{1}{3}(-(
\beta ^{4}(3-4\beta ^{2}+\beta ^{4}))^{\frac{1}{2}}
+|\beta|^{2})\},\nonumber\\
&&C(\rho_{AB_{\uppercase\expandafter{\romannumeral2}}})
=C(\rho_{AC_{\uppercase\expandafter{\romannumeral2}}})\nonumber\\
&&=2\max\{0,-\frac{1}{3}((\beta^{2}-1)(-1+(2p-1)r)(\beta^{2}+1+\nonumber\\
&&(p-1+\beta^{2}(2p-1))r ))^{\frac{1}{2}}\},
\frac{1}{3}|(1-\beta ^{2})^{\frac{1}{2}}(1-r)^{\frac{1}{2}}|-\frac{1}{3}\nonumber\\
&&(-(\beta^{2}-1)(1+(2p-1)r)(r-pr+\beta^{2}(1+r-2pr)))^{\frac{1}{2}},\nonumber\\
&&C(\rho_{B_{\uppercase\expandafter{\romannumeral1}} C_{\uppercase\expandafter{\romannumeral2}}})=C(\rho_{B_{\uppercase\expandafter{\romannumeral2}}  C_{\uppercase\expandafter{\romannumeral1}}})\nonumber\\
&&=2\max\{0,\frac{1}{3}
((2-3\beta ^{2}+\beta ^{4})(\beta ^{2}+\beta ^{4}))^{\frac{1}{2}},\nonumber\\
&&\frac{1}{3}(-(\beta ^{2}-2\beta ^{6}+\beta ^{8})^{\frac{1}{2}}+|\beta(1-\beta^{2})^{\frac{1}{2}}|)\},\nonumber\\
&&C(\rho_{B_{\uppercase\expandafter{\romannumeral2}} C_{\uppercase\expandafter{\romannumeral2}}})\nonumber\\
&&=2\max\{0,-\frac{1}{3}((\beta|^{4}-1)^{2})^{\frac{1}{2}},\nonumber\\
&&\frac{1}{3}(-(2\beta|^{2}-3\beta|^{4}+\beta|^{8})^{\frac{1}{2}}+|1-\beta|^{2}|),\nonumber\\
&&\beta=(e^{-8\pi\omega(M-\alpha)}+1)^{-\frac{1}{2}}.
\end{eqnarray}

\begin{figure}
\begin{center}
\includegraphics[width=6cm]{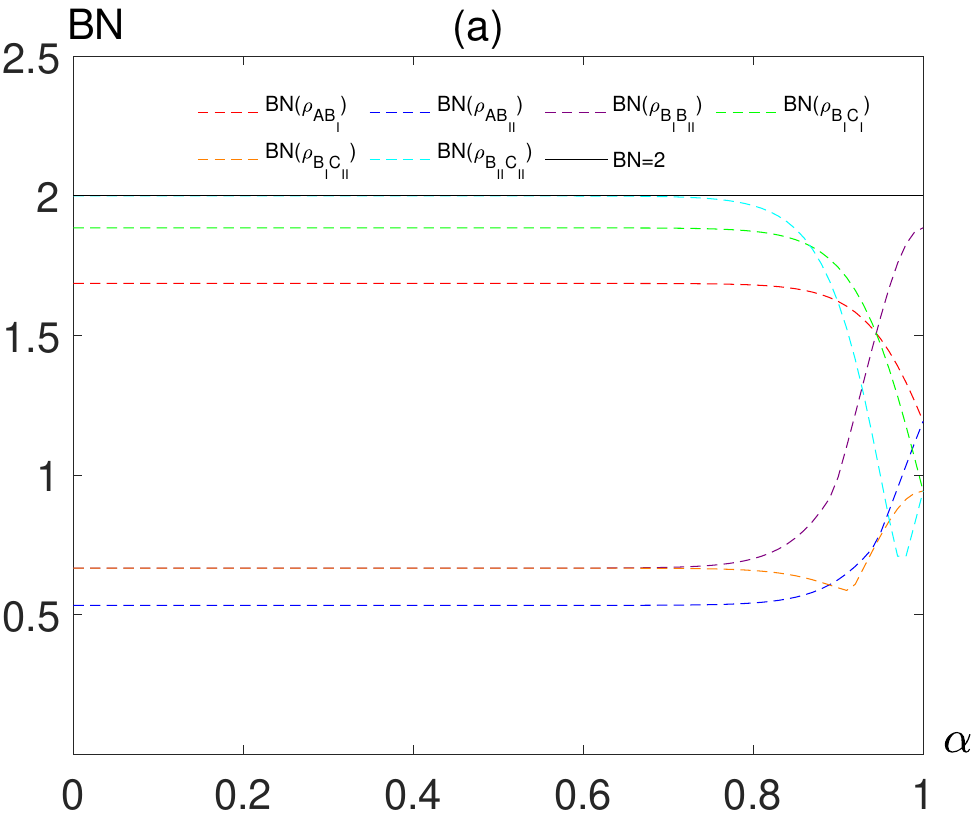}
\includegraphics[width=6cm]{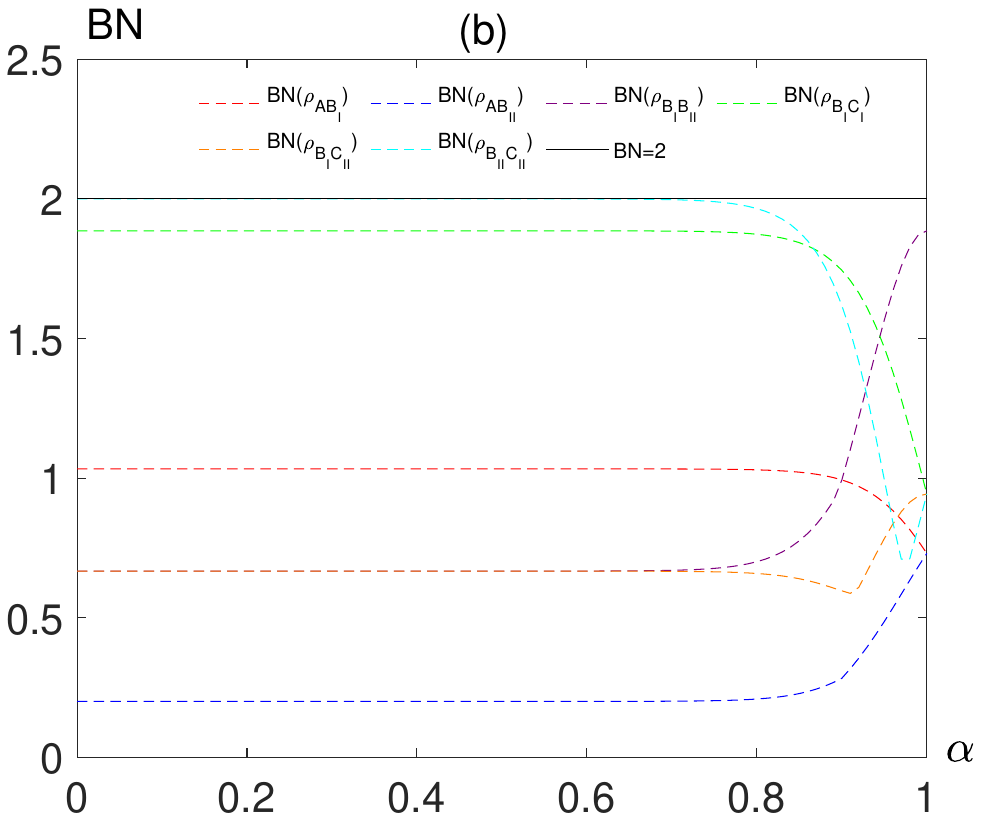}
\caption{\label{fig:fig9} 
 (a)  The BN is quantified by calculating the maximal violation of the CHSH inequality for different subsystems, with the parameters set as $\omega=1$, $M=1$, $r=0.2$ and $p=0.5$. (b) The BN is quantified by calculating the maximal violation of the CHSH inequality for different subsystems, with the parameters set as $\omega=1$, $M=1$, $r=0.7$ and $p=0.5$.}
\end{center}
\end{figure}

In Fig.9, we display the Bell nonlocality of all bipartite subsystems as a function of the dilaton parameter $\alpha$ for different decoherence strengths $r$. It is shown that irrespective of the value of $r$, no Bell nonlocality emerges in any bipartite subsystem. This result indicates that the combined effects of the black hole and environmental decoherence completely destroy the initial GTN in the physically accessible region, precluding its redistribution to any other tripartite or bipartite subsystems. Consequently, in the presence of decoherence and the dilaton black hole, quantum nonlocality can neither traverse the event horizon nor be transferred to any bipartite subsystems, thereby preventing its generation among these subsystems.

\begin{figure}
\begin{center}
\includegraphics[width=6cm]{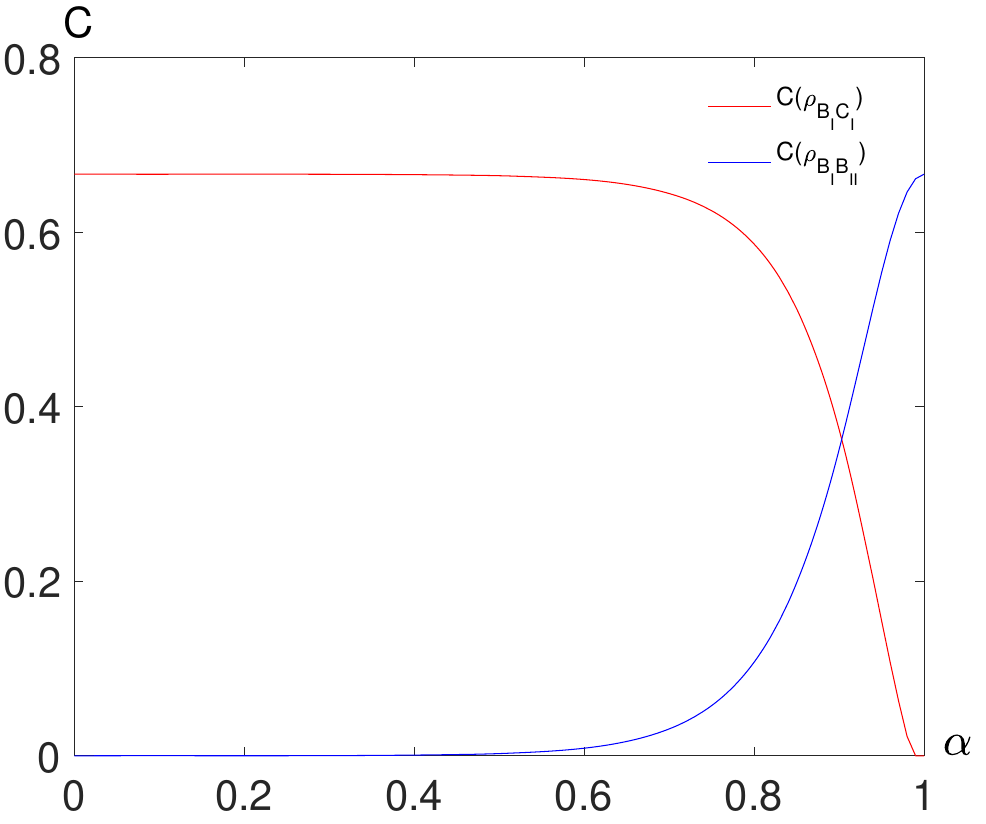}
\caption{\label{fig:fig10} 
  The concurrence for the reduced two-qubit subsystems $\rho_{B_IB_{II}}$ and $\rho_{B_IC_I}$ are plotted as a function of dilaton parameter $\alpha$ with $\omega=1$, $M=1$, $r=0.2$ and $p=0.5$.}
\end{center}
\end{figure}

In Fig.10, the entanglement $C(\rho_{B_{\uppercase\expandafter{\romannumeral1}} C_{\uppercase\expandafter{\romannumeral1}} })$ of the physically accessible region and the entanglement $C(\rho_{B_{\uppercase\expandafter{\romannumeral2}} C_{\uppercase\expandafter{\romannumeral2}} })$ of the physically inaccessible region are plotted as functions of the dilaton parameter $\alpha$. From the solid red line in the figure, it can be clearly seen that the entanglement $C(\rho_{B_{\uppercase\expandafter{\romannumeral1}} C_{\uppercase\expandafter{\romannumeral1}} })$ of the physically accessible region remains nearly constant during the initial evolution and then begins to decay once the dilaton parameter $\alpha$ becomes relatively large. This implies that the dilaton effect of the black hole drives quantum entanglement out of the physically accessible region and into the subsystems of the physically inaccessible region. In addition, the solid blue line further reveals that the entanglement $C(\rho_{B_{\uppercase\expandafter{\romannumeral2}} C_{\uppercase\expandafter{\romannumeral2}} })$ in the physically inaccessible region remains zero initially and then increases gradually. This behavior indicates that the dilaton effect of the black hole can generate the quantum entanglement in the physically inaccessible region. Since only Alice is subject to environmental decoherence, the entanglements $C(\rho_{B_{\uppercase\expandafter{\romannumeral1}} C_{\uppercase\expandafter{\romannumeral1}} })$ and $C(\rho_{B_{\uppercase\expandafter{\romannumeral2}} C_{\uppercase\expandafter{\romannumeral2}} })$ are unaffected by the decohering environment. These results therefore show that quantum entanglement of bipartite subsystems can be redistributed across the event horizon via the dilaton effect even in the presence of decoherence.

\section{CONCLUSION}
In this paper, we investigate how the dilaton effect of the GHS black hole affects quantum nonlocality and entanglement for a system initially prepared in the genuine tripartite entangled $\left | W \right \rangle $ state under decoherence. Using established numerical methods from the literature\textsuperscript{\cite{3x}}and the $\pi$-tangle measure, we analyze the dynamics of GTN and GTE in the tripartite subsystem. Our results show that both GTN and GTE in the physically accessible region decrease as the dilaton parameter $\alpha$ increases. In particular, GTN undergoes "sudden death" at a certain critical value of $\alpha$, whereas GTE remains constant at first and then decays to a small, nonzero value when $\alpha$ becomes sufficiently large. In addition, we find that both GTN and GTE decrease as the decoherence strength $r$ increases. Notably, when the decoherence strength $r$ is relatively large, the GTN in the physically accessible region vanishes entirely, whereas GTE persists, indicating that GTE exhibits stronger robustness against decoherence than GTN.

Furthermore, through the study of GTN and GTE in the physically inaccessible region, we find that although the GTN value remains nonzero for any dilaton parameter $\alpha$, it is always less than $4$. This bound implies that GTN cannot be generated in the physically inaccessible region. Consequently, GTN cannot cross the event horizon of the black hole and cannot be redistributed by the dilaton effect. By contrast, the behavior of GTE in the physically inaccessible region differs markedly from that of GTN. Specifically, as GTE in the physically accessible region decreases, GTE in the physically inaccessible region begins to emerge. This indicates that GTE can cross the event horizon and achieve redistribution of quantum entanglement. However, as the decoherence strength $r$ increases, both the growth amplitude and growth rate of GTE in the physically inaccessible region decrease, indicating that environmental decoherence suppresses the transfer of GTE from the accessible to the inaccessible region. Conversely, the growth amplitude and growth rate of GTE in the physically inaccessible region increase with the decoherence parameter $p$. Physically, $p$ represents the relative probability $p/(1-p)$ of excitation loss versus excitation gain in the quantum system under the GAD channel, therefore, a larger $p$ promotes greater outflow of GTE from the physically accessible region. For a fixed environmental decoherence strength $r$, more GTE from the physically accessible region enters the physically inaccessible region, thereby enhancing the GTE generated there.

Finally, we also investigate whether the genuine tripartite nonlocality in the physically accessible region transfers to its two-qubit subsystems to form bipartite nonlocality. The results show that regardless of the value of the decoherence strength $r$, no Bell nonlocality appears in any bipartite subsystem. This finding implies that quantum nonlocality not only fails to cross the event horizon but also does not transfer to any bipartite subsystem. In other words, the initial GTN of the system is completely destroyed by the combined effects of the dilaton black hole and the decohering environment. In addition, we investigate the quantum entanglement dynamics of arbitrary bipartite subsystems. It is found that while the dilaton effect and environmental decoherence lead to the vanishing of entanglement in most bipartite subsystems, a finite nonzero entanglement is preserved in several specific subsystems. Notably, as the dilaton parameter $\alpha$ increases, the bipartite entanglement $C(\rho_{B_{\text{I}} C_{\text{I}}})$ in the physically accessible region decreases, while the bipartite entanglement $C(\rho_{B_{\text{I}} B_{\text{II}}})$ in the physically inaccessible region increases. These trends  reveal that, under the combined influence of decoherence and the dilaton black hole, both tripartite and bipartite entanglement are redistributed from the physically accessible region to the physically inaccessible region across the event horizon.

\begin{acknowledgments}
This project was supported by the National Natural Science Foundation of China (Grant Nos. 12265007 and 11364006).
\end{acknowledgments}

\end{document}